**Invited review**

**Chondrule size and related physical properties: a compilation and evaluation of current data across all meteorite groups.**

Jon M. Friedrich[a,b,*], Michael K. Weisberg[b,c,d], Denton S. Ebel[b,d,e], Alison E. Biltz[f], Bernadette M. Corbett[f], Ivan V. Iotzov[f], Wajiha S. Khan[f], Matthew D. Wolman[f]

[a] *Department of Chemistry, Fordham University, Bronx, NY 10458 USA*
[b] *Department of Earth and Planetary Sciences, American Museum of Natural History, New York, NY 10024 USA*
[c] *Department of Physical Sciences, Kingsborough College of the City University of New York, Brooklyn, NY 11235, USA*
[d] *Graduate Center of the City University of New York, 365 5th Ave, New York, NY 10016 USA*
[e] *Lamont-Doherty Earth Observatory, Columbia University, Palisades, New York 10964 USA*
[f] *Fordham College at Rose Hill, Fordham University, Bronx, NY 10458 USA*

[*]Corresponding Author. Tel: +718 817 4446; fax: +718 817 4432.
E-mail address: friedrich@fordham.edu

**ABSTRACT**

The examination of the physical properties of chondrules has generally received less emphasis than other properties of meteorites such as their mineralogy, petrology, and chemical and isotopic compositions. Among the various physical properties of chondrules, chondrule size is especially important for the classification of chondrites into chemical groups, since each chemical group possesses a distinct size-frequency distribution of chondrules. Knowledge of the physical properties of chondrules is also vital for the development of astrophysical models for chondrule formation, and for understanding how to utilize asteroidal resources in space exploration. To examine our current knowledge of chondrule sizes, we have compiled and provide commentary on available chondrule dimension literature data. We include all chondrite chemical groups as well as the acapulcoite primitive achondrites, some of which contain relict chondrules. We also compile and review current literature data for other astrophysically-relevant physical properties (chondrule mass and density). Finally, we briefly examine some additional physical aspects of chondrules such as the frequencies of compound and "cratered" chondrules. A purpose of this compilation is to provide a useful resource for meteoriticists and astrophysicists alike.

## 1. Introduction

Early solid components of the Solar System included Calcium-Aluminum-rich Inclusions (CAIs), chondrules, and Fe-Ni metal and sulfide (primarily troilite, FeS) grains. The dimensions of each of those components generally fall in the μm to mm size range (Brearley and Jones, 1998; Ebel et al., submitted). Other silicate materials − materials that would become chondrite matrix − were also present, but their size ranges lie at the lesser end of or below the size distributions of the previously mentioned materials (Brearley and Jones, 1998; Ebel et al., submitted). Chondrules, or spherical objects of predominately silicate composition found in chondrites, contain essential information needed to elucidate chemical and astrophysical processes operating at the time of their formation during the early evolution of the Solar System. Numerous mechanisms for chondrule formation have been proposed, and there is general agreement that they formed from the rapid heating of predominantly silicate precursor materials followed by fast (10–1000 °C/hour) cooling (Hewins et al., 1996). Most chondrules are dominated by Fe- and Mg silicates in quenched silicate liquid (mesostasis), but many also contain reduced metal (Fe-Ni) and troilite (FeS). Chondrules typically make up between 20-80% of a chondrite by volume and their apparent diameters generally range from ~100 to ~2000 μm (Weisberg et al., 2006).

The diameters of chondrules provide a convenient criterion for chondrite classification and, more importantly, provide fundamental constraints necessary for the development and testing of astrophysical models for chondrule formation. Average chondrule sizes vary among (and possibly to a lesser extent, within) different chemical groups of chondrites, and the average apparent diameters of chondrules are considered a valid criterion for establishing the classification of a chondrite (Van Schmus and Wood, 1967; Weisberg et al., 2006). The size distributions of chondrules among the different chondrite groups could be a result of their mechanism of formation, a result of post-solidification nebular sorting, the result of a process on the parent body, or a combination of factors (e.g., Shu et al., 1996; Weidenschilling, 2000; Cuzzi et al., 2001; Cuzzi and Weidenschilling, 2006; Chiang and Youdin, 2010; Wurm et al., 2010). Whatever the case, the dimensions of chondrules provide substantive limits on their natal astrophysical environments.

In this work we compile historical data on the sizes and densities of chondrules. Chondrule dimensions (generally diameters measured in thin section − see Section 2) have been the most frequently reported. Dedicated studies of chondrule densities (a more difficult measurement) are sparser, but because of its astrophysical significance, we also compile literature data on density. Finally, we discuss the current knowledge and examine the prospects for future data refinement. One goal of this compilation is to provide a useful resource for meteoriticists, astrophysicists, and those contemplating exploration and exploitation of chondritic asteroids.

## 2. Notes on sources and compiled data

The majority of the data compiled and evaluated here are chondrule diameters, mainly apparent diameters measured on a two dimensional surface (i.e., petrographic thin sections). For ease of discussion and presentation of chondrule diameter data, we examine each chondrite chemical group separately and within each we proceed in order of publication date, from oldest to most recent so the evolution of data is apparent over time. We primarily consider data only on whole chondrules, but have occasionally included historically important data that included combined size data on the silicate grains and/or chondrule fragments which can be found in most chondrites (e.g., Stakheav et al., 1973; Dodd, 1976). Since chondrule diameter data based on studies that included chondrule fragments are inherently biased, when such data are included, it is noted in the narrative, Tables, and Figures. Petrographic studies on small numbers (n <~10) chondrules have generally been omitted because of the small sample size and specific (and generally unusual) chondrules studied (e.g., Krot and Rubin, 1994). Likewise, diameters for chondrules specifically isolated for isotopic or compositional studies have generally been excluded, because those studies also examined small numbers of chondrules and the sizes are biased because collecting instrumental data is easier with larger specimens. When such data are included, it too is noted in the narrative and Tables. At times, it was easy to extract non-tabular or graphical data for inclusion in our compilation. Other times it was more difficult. We only present data derived from graphical sources when we can do so with high confidence.

We do not address sizes of the fine grained rims that can be found on many chondrules, but again note in the narrative when such data are available. For example, Rubin (2010) addressed (igneous) rim sizes among different chemical classes of chondrites and Huang et al. (1996) list data on rim dimensions around LL chondrite chondrules.

Some investigators have reported chondrule sizes or distributions by type of chondrule petrographic texture (e.g., barred olivine, radial pyroxene, cryptocrystalline, etc.). In our compilations, we do not show data separated by chondrule type, but we do discuss this aspect later (Section 5.3.). In general, readers are referred to the original publications for detailed data and discussion.

We have not included data on crystalline lunar spherules (Symes et al., 1998), since it is unlikely that their origins are akin to those of chondrules. However, we do report on 10-100 μm microchondrules (e.g., Rubin et al., 1982; Bigolski et al., 2014) and macrochondrules and/or megachondrules (e.g., Weisberg et al., 1988a; Ruzicka et al., 1998; Weyrauch and Bischoff, 2012) in chondritic meteorites. However, the origins of some of these may be different than chondrules more typical in size.

Occasionally, in light of newer data and the resulting refinements in accepted chondrite classifications, chondrites have changed putative chemical group classifications. We always use the current chemical group classification for all chondrites. For example, Inman and Bjurböle are now generally considered intermediate L/LL chondrites and Bishunpur was once considered an L chondrite, but newer data indicate an LL classification is a better description. Other examples will be found below. Intermediate type H/L chondrites are described with the H chondrites. Because of the wealth of reported data for them, the L/LL chondrites (Inman and Bjurböle) are placed in their own section for ease of discussion and clarity of graphical presentation. However, we caution that it is unclear whether the intermediate L/LL ordinary chondrites unquestionably represent a separate chondritic parent body (and possibly a distinct astrophysical formation environment).

Most published chondrule size data to date rely on the measurement of chondrules in two-dimensional (2D) petrographic thin section. Because of this, the most reported measure of size is apparent diameter, rather than the potentially more astrophysically-relevant radius (radius is a factor in the calculation of the Stokes drag force on a spherical object; perhaps less relevant to chondrules, radius is also a factor in the quantification of particle drag in the Epstein regime). There are recognized issues with the determination of inherently three-dimensional (3D) parameters from 2D data (e.g., Chayes, 1956; Eisenhour, 1996; Higgins, 2006). However, numerous corrections, the most rigorous being based on the field of stereology, are available. Reviews of stereological corrections are abundant and the reader is encouraged to seek out the most recent (e.g., Mouton, 2011). Most chondrule size data are presented as apparent diameter, which is given without correction. However, some studies have made 2D to 3D corrections to the data. Dodd (1976) used the empirical conversion curves of Friedman (1958). Hughes (1978a) was the first to apply a theoretical numerical correction to measured mean and median apparent chondrule diameters, although it was undoubtedly incomplete (see Eisenhour, 1996). Some (e.g., Rubin and Keil, 1984) also used or referenced the Hughes (1978a) correction. Later, Eisenhour (1996) gave an improved means of correction and some (e.g., Kuebler et al., 1999) have implemented his means of correction. We comment on the validity of the well-regarded Eisenhour (1996) correction later (Section 7.2.1). In all cases of our compiled values, we point out in the narrative and Tables whether a 2D to 3D correction was attempted and which correction was applied.

The vast majority of chondrule diameter data in the literature has been statistically predigested, i.e., raw apparent diameter or similar dimensional data have been condensed into means, medians, ranges, and other descriptive statistics. We show these data in our figures. For studies involving the size description of many chondrules, data have usually been binned or are presented as discrete probability functions (histograms). Very few researchers report complete undigested data sets (i.e., a listing of all individual chondrule sizes examined during the study), but with the more recent possibility of electronic annexes we will suggest that authors report such data in the future. When possible, we graphically show these data and, for comparison among different studies, we present the data as diameter versus a normalized frequency. Although the correct and ideal means of displaying such data is with histograms, we use data points connected with lines to give an idea of the shape of an inferred probability density function. No data on the continuous probability density functions of chondrules have ever been presented in the literature. Some authors have used phi ($\varphi$) notation (Krumbein, 1936; Folk and Ward, 1957; cf. Folk, 1980), as used in sedimentology, for the description of chondrule sizes. Phi ($\varphi$) units are defined as [$\varphi = -\log_2$ diameter (mm)]. For ease of comparison and reference, we have converted these to more easily compared linear measures: we use exclusively the unit of micrometers ($\mu$m) or microns throughout the manuscript. Whenever possible we also use axes of the same scale between chondrule groups for ease of comparison. We note that some authors report chondrule dimension statistics assuming a normal or Gaussian distribution, but chondrule size-frequency data are clearly not normally distributed. Within early literature sources there was some debate about chondrule size distributions following Rosin's law (describing the cumulative distribution of particle sizes obtained by crushing brittle solid materials; Rosin and Rammler, 1933), a Weibull distribution, or log-normal distribution. Teitler et al. (2010) demonstrated that a Weibull or log-normal distribution does not accurately describe the continuous size-frequency distribution function of chondrules. We make no attempt to fit available data into a function as our primary goal is to provide a concise but complete compendium of current chondrule size data.

### 3. Chondrule diameters
#### 3.1. Ordinary chondrites
##### 3.1.1. H chondrites

Table 1 and Fig. 1 summarize known chondrule size (diameter) and distribution parameters within the H chondrites. In the earliest comprehensive study, Dodd (1976) reported data on diameters of silicate particles for eight H-group meteorites. This dataset included data on chondrule fragments [see Martin and Mills (1978) for clarification]. Dodd (1976) reported a median silicate particle diameter of 330 µm in H chondrites, but this included two chondrites (Tieschitz H/L3.6, Bremervörde H/L3.9) later recognized as being intermediate H/L chondrites. Removing these from consideration yields a median size of 320 ± 50 µm for silicate particles in H chondrites, which is similar to his 330 µm reported value. For interested readers, he also presented size distribution statistical parameters (based on the descriptive statistics of Inman, 1952) for the silicate grains.

Martin and Mills (1978) studied 1256 physically separated chondrules from the friable Allegan H5 chondrite. They found the mean diameter of these chondrules to be 570 µm with a median of 600 µm (Table 1; Fig. 1). The minimum chondrule diameter measured was 150 µm, and they state that care was taken to include the smaller size range chondrules in the study. A histogram of Allegan chondrule diameters shows a rapid increase from smaller values to the mean and then a gradual decline in diameter to the largest (2750 µm) chondrule measured (Fig. 1). Later, Martin and Hughes (1980) used this Allegan data and the data of others (Hughes 1978a; Stakheev et al., 1973) to compare mass frequency distributions between varieties of ordinary chondrites (OCs). However, since Allegan is an H5 chondrite, the data may be biased toward the chondrules that survived thermal metamorphism.

In an abstract, Gooding et al. (1978) reported the size, shape, mass, and density for 65 chondrules physically separated from an assortment of H3 and H4 chondrites. Their obviously size-biased selection yielded a diameter range of 900-1030 µm. Presumably, these chondrules were a subset of those used for the Gooding et al. (1980) compositional study. The same research group (Lux et al., 1981) examined correlations between compositions and textures of chondrules from several (then) putative unequilibrated H chondrites (but included the now recognized intermediate H/L Tieschitz and Bremervörde). Nonetheless, a weighted mean diameter of 420 µm can be calculated for the chondrules included in their suite. We do not plot the collective size distribution data of Gooding and Keil (1981), since it includes the H/L chondrite Tieschitz, but a histogram is available in the original publication. They presented some of the first data on sizes of chondrules as a function of chondrule type. Later, additional studies (Gooding and Keil, 1981; Gooding, 1983) presented more complete data on OC chondrule textural type by size. Although the data are biased towards larger chondrule sizes due to the compositional goals of the study, we show Gooding's (1983) data for H chondrites in Table 1. Goswami (1984) also opined on the size-frequency distributions of chondrule textural types within the H chondrites.

King and King (1979) examined 11 different OCs, including 6 then classified as H chondrites (Table 1) to determine the size frequency distributions of their chondrules. They studied whole chondrules only. They reported their data using the statistical parameters commonly used in sedimentology (Folk, 1980), which we have summarized in terms of median µm diameter in Table 1. From their work, they concluded that H chondrites have the smallest chondrule sizes among the OCs and that OCs possess a coarser chondrule size than chondrules found in CM or CO chondrites (see Section 3.5 and subsections on CM and CO chondrites below).

In their review, Grossman et al. (1988a) cited unpublished data along with those of King and King (1979), to estimate the mean diameter of H chondrite chondrules as 300 µm. Rubin (2000, 2005, 2010) and Weisberg et al. (2006) cited Grossman et al. (1988a) to quote a mean chondrule diameter of 300 µm in the H chondrites.

Kuebler et al. (1999) presented the first H chondrite chondrule diameters that utilized the stereological correction of Eisenhour (1996). To display the changes in the distribution due to the applied correction, Fig. 1 shows both the corrected size distribution, which yields a mean diameter of 460 µm (Table 1), and the uncorrected distribution. Statistical parameters for the uncorrected data were not given in the original publication.

In a figure, Cuzzi et al. (2001) illustrated Stokes number distributions for two H chondrites [Outpost Nunatak (OTT) A80301, H3.8; Grosvenor Mountains (GRO) 95524, H5], some of which were later summarized and presented by Teitler et al. (2010). They presented statistical summaries for Queen Alexandra Range (QUE) 93030 (H3.6) and GRO 95524 (H5). However, Teitler et al. (2010) distinguished between "picking" (QUE 93030 and GRO 95524) and disaggregating (GRO 95524 only) for the isolation of their chondrules for size and mass measurements. They concluded that their picked suites of chondrules were biased with respect to size, because the chondrules were simply picked from available fine material rather than systematically disaggregated. Given their conclusion, we summarize their most accurate (disaggregated) data for GRO 95524 in Table 1 and Fig. 1.

### 3.1.2. L chondrites

Table 2 and Fig. 2 summarize known chondrule size (diameter) and distribution parameters within the L chondrites. Stakheav et al. (1973; but cf. Lang et al., 1975) performed a disaggregation study to examine the size- and mass-frequency distributions of chondrules in three L chondrites: Elenovka (L5), Nikolskoe (L4), and Saratov (L4). They presented size-frequency data in tabular and graphical form but did not provide statistical summaries. Their size-frequency distributions are shown in Fig. 2. However, they included chondrule fragments in their compilations. Hughes (1980) used the Stakheav et al. (1973) data to discuss a possible relationship between chondrule size and bulk density, and Martin and Hughes (1980) used the data for a study on the mass frequency distribution of chondrules. The notebooks and raw data behind Stakheav et al.'s (1973) work no longer exist (M. Ivanova, personal communication, May 2014).

Dodd (1976) reported the diameters for "silicate particles", including chondrules, from six L chondrites (Table 2). This included data on chondrule fragments. He cited a median diameter of 460 micrometers. He also presented size distribution statistical parameters of Inman (1952) (sometimes used for terrestrial sediment grain sizes) for the silicate grains.

Gooding et al. (1978) reported diameters of 56 hand-picked chondrules from the L chondrite group (probably including at least 24 from Hallingeberg (L3.4) – see Gooding et al., 1980) and found that their mean apparent diameter, although size-biased (from the hand picking), was 1020 µm. They also determined the percentage abundances of chondrule types from the L chondrite group and concluded that chondrule size and shape are not strictly correlated with chondrule textural type. Since their chondrules were not representative, we do not plot their data here.

King and King (1979) examined the size frequency distributions of "fluid drop" (presumably round and whole) chondrules of two unequilibrated L chondrites. They used petrographic thin section measurements to determine the median apparent diameters and made no stereological correction. The 132 chondrules from Khohar (L3.6) were found to have a median apparent diameter of 620 µm (Table 2). The 58 chondrules from Mezö-Madaras (L3.7) were found to have a median apparent diameter of 490 µm. As with the H chondrites, King and King (1979) reported their data using the statistical parameters commonly used in sedimentology, and we summarize available data in Table 2.

Ikeda and Takeda (1979) conducted a petrographic examination of Yamato-74191 (Y-74191), L3.7, with a focus on bulk chondrule compositions. The apparent sizes of different groups of nearly-round chondrules larger than 200 μm in diameter were measured under a microscope. The frequency of chondrule sizes was shown in the original publication, but no numerical values were given. The mean diameters of the chondrules are shown in Fig. 2. The range of chondrule diameters inferred from their graphical representation is 200-2000 μm (Table 2).

Nagahara (1981) conducted a petrographic study of chondrules in the L3.5 Allan Hills A77015 (ALH A77015) to investigate a correlation between size, bulk chemical composition, and texture of 108 chondrules. Measurements of size were made on thin sections and the average apparent diameter was estimated as ~800 μm. She noted no relation between chondrule size and texture. Nagahara (1981) provides a plot of size distribution with respect to textural type, and we reproduce the size frequency distribution including all chondrules in Fig. 2. There is probably some bias towards larger chondrule sizes in this dataset because of the compositional goals of this study.

Gooding and Keil (1981) reported frequencies of chondrule textural types within L chondrites. They also reported the first data on compound and cratered chondrules, which may place constraints on chondrule collisions within a nebular environment. They reported upper limits for the frequency of compound chondrules as ~4% and cratered chondrules as 10%. We do not plot the size distribution data of Gooding and Keil (1981), since the data for L chondrites contain the now recognized LL chondrite Bishunpur in the summary.

Gooding (1983) reported size, mass, and density of chondrules in three different L chondrites. He examined Hallingeberg (L3.4), Saratov (L4), and Tennasilm (L4) (Table 2), measuring the maximum dimensions of 8 to 26 chondrules (the actual number is unclear) and reporting the geometric means (Table 2). A size bias toward larger chondrules is evident in the data (Table 2), and this was noted by the author.

Rubin and Keil (1984) measured the size range and mean of chondrules within the L3.5 chondrite ALH A77011. They identified 163 barred olivine chondrules with a mean diameter of 625 μm and 70 radial pyroxene and cryptocrystalline chondrules with a mean diameter of 622 μm. Later Rubin and Grossman (1987) cited a mean value of 476 μm for all chondrules in this chondrite from unpublished data from the Rubin and Keil (1984) study.

Grossman et al. (1988a) reviewed both the physical, textural, and chemical properties of chondrules. They cited the data of Rubin and Keil (1984 and unpublished data) and King and King (1979) to arrive at their assessment that the mean L chondrite chondrule diameter lies between 600 and 800 μm. Weisberg et al. (2006) cited 700 μm from the Grossman et al. (1988a) work as a mean chondrule diameter. Rubin (2000, 2005) cited Grossman et al. (1988a) but quoted a mean apparent diameter of 500 μm. Later, Rubin (2010) quoted 400 μm as an estimate for an L chondrite chondrule mean diameter.

In an abstract, Paque and Cuzzi (1997) reported the mean diameter of chondrules in ALH 85033 (L4) as 720 μm (Table 2, Fig. 2). Cuzzi et al. (1999) mentioned that the work was based on 235 chondrules. The measurements were made by disaggregation and masses (and densities) were also measured. Cuzzi et al. (2001) give a graphical distribution of the Stokes number for this set of chondrules. Teitler et al. (2010) expanded on these data, conducting an examination of statistical tests to determine the nature of chondrule sorting. They both disaggregated and "picked" the chondrules from ALH 85033. Since they found that the "picked" chondrules may have some size bias, we summarize the data only for the disaggregated chondrules in ALH 85033 (Table 2, Fig. 2). We note that they also determined the masses (and densities) of chondrules in ALH 85033.

Metzler (2012) examined an unequilibrated (<3.5) L chondrite clast (described in Metzler et al. 2011) in the Northwest Africa (NWA) 869 L3-6 chondrite. He measured the diameters of 67 chondrules and found their mean diameter to be 520 μm.

### 3.1.3. Bjurböle and Inman, L/LL chondrites

Although the L and LL chondrites possess relatively distinct olivine compositions and kamacite Co contents, some chondrites rest between definite compositional cutoffs (Kallemeyn et al., 1989). These are sometimes presented as an intermediate group: the L/LL chondrites. An example of this is Bjurböle (L/LL4), which also happens to be exceptionally friable and, hence, its chondrules are often studied because they are easily physically separated. We summarize Bjurböle and Inman (also acknowledged as an L/LL chondrite, classified as L/LL3.4) chondrule diameter data in Table 3 and Fig. 3, and here present a narrative of efforts to measure chondrule sizes in them.

Stakheav et al. (1973) and a follow-up abstract (Lang et al., 1975) performed the first of several disaggregation studies to examine the size- and mass-frequency distributions of chondrules in Bjurböle. They presented size-frequency data in tabular and graphical form but did not provide statistical summaries. We show their data in Fig. 3. Stakheav et al. (1973) concluded that the frequency of chondrules increases with decreasing size (Fig. 3); however, Martin and Hughes (1980) reported that both whole chondrules and chondrule fragments along with silicate particles were included in the Stakheav et al. (1973) study, so some caution is recommended when comparing their results with those of others. A minimum chondrule size may exist. Nevertheless, Martin and Hughes (1980) use the Stakheav et al. (1973) data to fit a Weibull function.

Dodd (1976) performed thin section measurements of silicate particles in Bjurböle and presented median values (Table 3, Fig. 3), but, as noted above (see Section on H chondrites) he did include all silicate particles in addition to whole chondrules, so the median diameter value is biased toward the low end.

Martin and Mills (1976) measured 97 separated chondrules from Bjurböle and reported histograms and common statistical parameters (Table 3). They found a mean of 1180 μm for Bjurböle chondrule diameter. They maintained that their lower limit of 400 μm was real and not an experimental artifact, but given the differences between their data and the data of others (Fig. 3) there was almost certainly a size bias.

Hughes (1977, 1978a) reported results from a combined disaggregation and thin section analysis of the size distributions of chondrules within Bjurböle (Table 3, Fig. 3). He only included clearly spherical chondrules in the disaggregation study, so some bias may be present. He also used this and additional data (Hughes, 1978b, 1980) to examine interrelationships between chondrule diameter, mass, and density. Hughes (1978b) presented a histogram of his disaggregated Bjurböle chondrule sizes (Fig. 3). Paque and Cuzzi (1997) later cited the Hughes data.

Studying whole chondrules only, King and King (1979) evaluated the chondrule size distribution in the Inman chondrite. The reported their data using the statistical parameters commonly used in sedimentology (Folk, 1980), which we have summarized in terms of median μm diameter in Table 3 and Fig. 3.

Rubin and Keil (1984) examined 374 chondrules in the unequilibrated Inman chondrite and reported their results separated into two chondrule textural type groupings (barred olivine and radial pyroxene plus cryptocrystalline). They reported typical statistical descriptors, which we summarize in Table 3. The histogram of frequency versus binned diameters for Inman for both types of chondrules is shown in Fig. 3. Later, Rubin and Grossman (1987) cited a mean value of 688 μm for all chondrules in this chondrite from unpublished data from the Rubin and Keil (1984) study (Table 3).

Kuebler et al. (1999, also see Kuebler et al., 1997) presented chondrule size distributions (Fig. 3) and statistical parameters for chondrules measured in thin section, corrected for bias by the Eisenhour (1996) numerical technique in graphical form. We show the corrected data in Fig. 3; to compare uncorrected and corrected distributions, we refer the reader to the H chondrite data above (see Fig. 1 for a comparison).

Cuzzi el al. (2001) presented Stokes parameter number distributions of 150 chondrules separated from the Bjurböle chondrite. Later, Teitler et al. (2010) presented a statistical summary of these chondrules, but noted that the Bjurböle data were "picked" rather than disaggregated, so some bias is probable (see H and L chondrite sections for clarification).

*3.1.4. LL chondrites*

Table 4 and Fig. 4 summarize known chondrule size statistical descriptors and distribution parameters within the LL chondrites. The first study of LL chondrite chondrule sizes was done by Dodd (1976). He reported data on diameters of "silicate particles" for six LL individuals (Table 4). Various statistical parameters describing the size frequency data of the silicate particles were presented. We show the median values in Table 4 and Fig. 4. For interested readers, the same statistical descriptions were given for metal particles in the LL chondrites. Martin and Mills (1976) extracted 245 chondrules from the Chainpur (LL3.4) chondrite by gentle crushing and hand-picking. They measured these individual chondrules using binocular microscopes to investigate their size distribution and shape. They report the mean diameter as 1090 μm and the median as 1020 μm (Table 4). Like Dodd (1976), they provide statistical summaries using the conventions of sedimentary petrology (in this case based on Inman, 1952 and Krumbein and Pettijohn, 1938). A graphical summary of their size distribution data is shown in Fig. 4.

Hughes (1978a) also examined LL chondrules within the Chainpur meteorite using petrographic thin sections. For chondrules examined by petrographic thin sections corrected to true values, he reported a mean of 893μm and median of 817μm (Table 4).

In an abstract, Gooding et al. (1978) report a mean diameter of 1280 μm for 70 LL chondrite chondrules separated from their parent meteorites. The meteorites from which the probably size-biased chondrules came from are not listed.

King and King (1979) studied the size frequencies of 45 LL3 chondrules from the Parnallee and Bishunpur meteorites using petrographic thin sections. They reported a median diameter of 366 μm and 637 μm, respectively, and examined only "fluid drop", or round, chondrules. They summarized the grain size statistics using the manner of Folk (1980).

Gooding and Keil (1981) presented data on LL chondrule textural type by size. Interested readers are pointed to this work for further information. Although the data are seemingly biased towards larger chondrule sizes (see Fig. 4), we show their LL chondrite chondrule size distribution data in Table 4. They point out that LL chondrite chondrules are, on average, larger than chondrules in the L and H chondrites. Gooding (1983) reported the geometric means (Table

4) of chondrules in each of two thin sections from 4 different LL chondrites as well as expanding on the abundances of textural types of chondrules. He acknowledges the likelihood of a size selection bias in the Gooding (1983) and all his previous works. Because of the small sample size and probable bias we do not show the diameters graphically.

Grossman et al. (1988a) reported that a best estimate for the mean LL chondrite diameter is 900 µm. Weisberg et al. (2006) referred to this value for chondrule sizes in LL chondrites.

Huang et al. (1996) measured the diameters of chondrules in Semarkona (LL3.0) and Krymka (LL3.2). We summarize their reported results for individual chondrules in Table 4 and show a histogram of their data in Fig. 4. Huang et al. (1996) were also among the first to report on the thickness of the fine grained rims on chondrules.

Kuebler et al. (1999) presented data for chondrule diameters from the Kelly LL4 chondrite. They utilized 2D to 3D stereological corrections (Eisenhour, 1996) in the presentation of their petrographic thin section measurements. Fig. 4 shows the corrected size distribution, which possesses a mean diameter of 660 µm.

Nelson and Rubin (1999, 2002) measured apparent diameters of chondrules from several LL chondrites. In 1999, they reported on 236 LL chondrules from Semarkona with the average (mean) to be 560 µm. They compared this to the 900 µm mean value determined for LL chondrules reported by Grossman et al. (1988a), noting that their value is noticeable smaller. Continuing their work, Nelson and Rubin (2002) reported size distributions for five unequilibrated LL chondrites. They measured a total of 719 intact chondrules from Semarkona (LL3.0), Bishunpur (LL3.15), Krymka (LL3.2), Piancaldoli (LL3.4), and Lewis Cliff (LEW) 88175 (L3.4) and reported the mean apparent diameter of the total 719 intact chondrules to be 570 µm. Readers are referred to Table 4 for mean diameters of the LL chondrules from each meteorite and Fig. 4 for a graphical summary. We note here (but also see Section 6.1) that Nelson and Rubin also examined the size frequency distribution of chondrule fragments. They inferred that different chondrule textural types are more easily fragmented than others by impacts on the parent body, leading to skewed distributions of size frequency distributions of different textural types. Rubin (2005, 2010) used these newer, more complete, data on whole chondrules when reporting LL chondrite chondrule sizes.

The most recent study of apparent diameters of LL chondrules can be found in Metzler (2012). He conducted a study of chondrule textural types and their mean degree of deformation in "cluster" chondrite clasts and clastic meteorite fragments and reported the mean apparent diameter of the LL chondrules (calculated from measured chondrule cut faces in thin sections) from five meteorites as reported in Table 4. We note that most of the chondrite clasts studied by him may not be representative of LL chondrites as a whole; however, they are of LL chondrite composition.

### 3.2. Enstatite chondrites
### 3.2.1. EH chondrites

Rubin and Grossman (1987) separated 63 chondrules from the Qingzhen EH3 chondrite. A histogram of their sizes is shown in Fig. 5. They acknowledged that they likely omitted smaller chondrules from their disaggregation study. However, they gave size distribution data for chondrules of different textures and, based on petrographic thin section measurements of 689 chondrules in Qingzhen (EH3), Kota-Kota (EH3), and ALH A77156 (EH3) they found a mean diameter of 213 µm. They also give detailed information on the size distributions of chondrules by textural type for each of the chondrites investigated. Rubin (2000, 2010) cited this work and quoted a mean EH chondrule size of 220 µm.

Schneider et al. (2002, but also see Schneider et al., 1998) gave ranges for EH chondrule diameters and a thorough breakdown of textural types by size. They found a mean chondrule diameter of 278 µm in three EH chondrites. We show a histogram of their data in Fig. 5.

Weisberg et al. (2011) found some chondrules in the EH chondrites Sahara 97096 and Yamato 691 in the range of 500-1000 µm and one barred olivine chondrule reaching >3000µm in diameter. This study excluded smaller chondrules.

### 3.2.2. EL chondrites

Rubin (2000) cited unpublished work and Rubin et al. (1997) to give a mean chondrule size of 550 µm for EL chondrites. The values quoted by Rubin (2010) were derived from the Rubin (2000) work. Schneider et al. (2002) gave ranges for EL chondrule diameters and a breakdown of the size distributions of textural types. They found a mean chondrule diameter of 476 µm in three EL chondrites (Table 5). We show a histogram of their EL chondrule data in Fig. 5.

### 3.3. R chondrites

The first reports of chondrule sizes in the R chondrites were based on Carlisle Lakes (R3.8) and ALH 85151 (R3.6), which suggested their mean diameter lies between 400-500 µm (Rubin and Kallemeyn, 1989). Kallemeyn et al. (1996) reported more complete chondrule diameters for a variety of R chondrites. Data are shown in Table 5 and graphically in Fig. 5. They reported a mean R chondrite chondrule diameter of 400 µm, based on the measurement of 7 chondrites. Rubin (2000, 2010) cited the Kallemeyn et al. (1996) data and quoted the 400 µm chondrule diameter as the best value for the R chondrites.

### 3.4. K chondrites

Weisberg et al. (1996) examined three putative K chondrites and confirmed that they are an independent chemical group. They measured the chondrule diameters in them (Table 5, Fig. 5) but did not give a definitive mean group value. However, Weisberg et al. (2006) suggested a mean apparent chondrule diameter of ~600 µm for the group. Genge and Grady (1999) reported on the abundances of chondrule textural types (see Section 6.1) and described the chondrule rims in the Kakangari K chondrite.

### 3.5. Carbonaceous chondrites
### 3.5.1. CM chondrites

Rubin and Wasson (1986) discussed the compositional differences between CM and CO chondrites. They found that one hundred chondrules from the Murray CM2 chondrite had a mean diameter of 270 µm. Other studies (Grossman et al., 1988a; Weisberg et al., 2006) have cited this work as the basis for the 300 µm suggested for chondrule sizes in CM chondrites.

### 3.5.2. CO chondrites

King and King (1978) reported "silicate grain" sizes from a study of five CO chondrites. However, they do not give statistical parameters for whole chondrules, so we set this work aside for the CO chondrites as well as the rest of the carbonaceous chondrites examined by King and King (1978). In a personal communication cited by Rubin and Wasson (1988), King and King found a mean diameter of 196 µm for CO chondrules. It is unreported how many chondrules were measured to arrive at this value.

Rubin (1989a) reported on the size frequency distribution of chondrules in CO chondrites. We show the mean and standard deviation he found (148 +132/ -70 µm) after examining a total of 2834 CO chondrite chondrules (Table 6, Fig. 6). For interested readers, Rubin (1989a) shows size distributions for 11 individual CO chondrites along with a breakdown of textural information by size. This is and remains the largest number of chondrules examined in a single study to date. Finally, Rubin (1998) used this and some additional data to examine a possible relationship between petrographic type and chondrule diameter in CO chondrites. Rubin (2000, 2010) also cites the Rubin (1989a) work for CO chondrule sizes. Eisenhour (1996) used the Rubin (1989a) dataset as a test for a stereological correction for petrographic section based measurements (Fig. 6).

May et al. (1999) measured chondrule diameters in the Warrenton (CO3.7), Lancé (CO3.5), and ALH A77307 (CO3.0) meteorites, finding average diameters of 259, 297, and 289 µm respectively. It is unclear what number of chondrules was studied to arrive at these means. Their values fall at the higher end of those found by the Rubin (1989a) and King and King (1978) studies. Finally, Moggi-Cecchi et al. (2006) reported a mean diameter of 110 µm for chondrules in the Acfer 374 CO3 chondrite.

### 3.5.3. CV and CK chondrites

Some evidence exists (Greenwood et al., 2010; Wasson et al., 2013) that the CV and CK chondrites are genetically related (i.e., their chondrule dimensions may be identical), but we consider them individually below.

### 3.5.3.1. CV chondrites

CV chondrites consist of three chemical subtypes: $CV_{ox-B}$, $CV_{ox-A}$, and $CV_{red}$ (Weisberg et al., 1997). Each of these chondrites probably formed from the same batch of nebular material and their chondrules probably experienced similar chondrule forming environments differing in availability of water (Ebel et al., submitted) (i.e., their chondrule dimensions should be identical and will be considered here as one group). McSween (1977) noted that individual chondrules in CV chondrites range from 550-2000 µm in diameter. The Grossman et al. (1988a) compilation estimated the chondrule diameter of CV chondrites to be 1000 µm. They did not cite the data sources used to arrive at this figure; nevertheless, Weisberg et al. (2006) used this as a best estimate for chondrule diameters in CV chondrites.

In an abstract, May et al. (1999) reported that the CV chondrules in Vigarano (CV3), Efremovka (CV3), Mokoia (CV3), and Leoville (CV3) had respective average (mean) diameters of 680, 655, 683, and 823 µm. Paque and Cuzzi (1997) disaggregated chondrules from Allende (CV3) and found a mean diameter of 850 µm. Rubin (2000) and Wasson et al. (2013) cited this work and reported a mean diameter of 910 µm for chondrule diameters in CV chondrites. Teitler et al. (2010) gave mean, median, and range data on the radii of chondrules in several CV chondrites. Unlike the other studies, they obtained their data from large numbers of chondrules disaggregated from larger samples (Table 6, Fig. 6).

### 3.5.3.2. CK chondrites

Kallemeyn et al. (1991) reported the apparent mean diameter of chondrules in the CK meteorites to range from 500 to 750 um. They also reported that the few discernable chondrules in the more recrystallized LEW 86258 (CK4) and Pecora Escarpment (PCA) 82500 (CK4/5) have diameters of 700 and 1000 μm respectively. Geiger et al. (1993) reported a mean and standard deviation of 870±380 μm in the anomalous CK3 chondrite Watson 002. Zipfel et al. (2000) found a range of 200-1000 μm for chondrules in Dar al Gani (DaG) 431, another anomalous CK chondrite. Tomeoka et al. (2003) reported a mean chondrule diameter of 750 μm in Kobe. Moggi-Cecchi et al. (2006) – also see Prastesi et al. (2006) for an initial report – found that the mean diameter of chondrules in Hammadah al Hamra (HaH) 337, a CK4 chondrite, is 700 μm. Rubin (2010) examined the NWA 1559 CV chondrite and found a mean chondrule diameter of 890 μm. Wasson et al. (2013) used the median diameter (870 μm) that Rubin (2010) found as evidence that the CK chondrites are related to the CV chondrites and concluded that chondrites of the CV3 and CK3 groups have indistinguishable mean diameters of 910 and 870 μm respectively.

### 3.5.4. CR chondrites

In an initial report, Bischoff et al. (1993a) noted the mean diameter of several CR chondrite chondrules to be 1000 ± 600 μm (Fig. 6, Table 6). Kallemeyn et al. (1994) collected chondrule diameter data on five CR chondrites (Table 6). They report the apparent diameters and size distributions of the chondrules (Table 6, Fig. 6). They found the mean diameters of CR chondrules to be 700 μm, which Rubin (2000) uses as a best mean for CR chondrites. However, it is important to note that some CR chondrules are complex multilayered objects with igneous rims (e.g., Weisberg and Prinz, 1991; Weisberg et al., 1993; Ebel et al., 2008). It is not clear that all authors use the same delineation for the chondrule edge.

### 3.5.5. CH chondrites

CH chondrites are composed of chondrules, metal and other inclusions but lack interstitial matrix material. They have unusual characteristics including the lack of matrix, dominance of (relatively small) cryptocrystalline chondrules, and a high abundance (~ 20 vol%) of FeNi-metal (e.g., Grossman et al., 1988a; Scott; 1988; Weisberg et al., 1988b; Bischoff et al., 1993b). The unusual characteristics of these chondrites led to the interpretation that their chondrules formed as a result of an asteroidal collision and are not truly primary materials formed in the solar nebula, as is proposed for chondrules in other chondrite groups (e.g., Wasson and Kallemeyn, 1990).

For chondrule diameters in the ALH 85085 CH chondrite, Scott (1988) reported a mean of 20 μm with a range of <4 to 200 μm. Grossman et al. (1988b) and Weisberg et al. (1988b) gave a similar estimate of 20 μm as the typical size for CH chondrules. Grossman et al. (1988b) also reported chondrule sizes by textural type. Wasson and Kallemeyn (1990) remarked on the small chondrule dimensions in ALH 85085. From measurement of 202 chondrules, Bischoff et al. (1993b) found the mean chondrule diameter for the Acfer 182 CH (and paired samples Acfer 207 and 214) to be 90 μm ±60 with the largest chondrule being 1100 μm. Based on a study of 170 chondrules from the Acfer 366 CH chondrite, Moggi-Cecchi et al. (2006) derived a mean chondrule size of 110 μm with a range of 35 to 450 μm.

Ischeyevo is a CH-CB breccia containing metal-rich and metal-poor lithologies (Ivanova et al., 2008). They reported that the metal-rich lithologies typically contain smaller chondrules with an average size of 100 μm (range is 20 to 400 μm), whereas the metal-poor lithologies have an average chondrule size of 400 μm with a range of 100 to 1000 μm. The latter chondrules may be more closely related to those in CB chondrules.

### 3.5.6. CB chondrites

The CB chondrites are another group of metal-rich chondrites with up to 80 % metal by thin section area (Weisberg et al., 2001). They are divided into CBa and CBb subgroups based on their metal abundances and sizes of their components. The CBa chondrites have chondrule-like objects up to one cm in size, whereas the CBb chondrites have chondrules up to about 1 mm with most about 200 μm in size (Weisberg et al., 2001). Weisberg et al. (2006) report a chondrule diameter of 200-1000 μm in the CB chondrites.

### 3.6. Grouplets, ungrouped, and anomalous chondrites

Many individual chondrites exist that cannot be unequivocally placed within the established groups covered in detail above. In other cases, marginal numbers of related chondrites exist (forming a grouplet). We mention these unusual cases here, but because of their odd nature, we do not show the data graphically or in compiled tables. The apparently unique carbonaceous chondrite LEW 85332 has a mean apparent chondrule diameter of 170 μm (Rubin and Kallemeyn, 1990). Chondrules in the chondritic clasts in the Netschaëvo iron meteorite (Bild and Wasson, 1977) range in apparent diameter from 300-1200 μm with a mean and standard deviation of 720 +360/-240. (Rubin, 1990). Kallemeyn and Rubin (1995) discussed the Coolidge and Loongana 001 chondrites, which seem to be a chemically distinct chondrite grouplet. They found that chondrules had an apparent diameter ranging from 190 μm to 2900 μm and an average of 700 (+930/-400) μm. Wang and Hsu (2009) reported the apparent diameter of chondrules in the unique carbonaceous chondrite Ningqiang as ~550 μm, based on 122 chondrules. Konrad et al. (2010) reported mean apparent chondrule diameters (n=593) of 70 μm in the ungrouped carbonaceous chondrite Acfer 094. Choe et al. (2010) examined the chemical and petrologic properties of 15 individual unusual carbonaceous chondrites, providing size ranges and mean diameters for inclusions in each of them.

### 3.7. Primitive achondrites (acapulcoites)

Primitive achondrites are the partial melt residues of chondritic precursors that have been subjected to different degrees of partial melting. Because of the high degree of recrystallization, they do not typically contain chondrules; however, some relict chondrules have been noted within the acapulcoite primitive achondrites. Yanai and Kojima (1991) found a ~250 μm barred olivine chondrule relict and McCoy et al. (1996) found a 1300×1900 μm sized relict chondrule in Monument Draw. Rubin (2007) found relict chondrules in the acapulcoite Dhofar 1222 to be ~700μm in mean apparent diameter, with a range of 300-1400 μm. Graves Nunataks (GRA) 98028 also contains relict chondrules that are 400 - 700 μm in diameter (Rubin 2007).

## 4. Chondrule bulk density

There have been some measurements of chondrule densities, but many have relied on estimates since disaggregation studies are the only means of obtaining both a volume and a mass. For example, Kuebler et al. (1997, 1999) reported the density of their chondrules but used assumptions about the chondrule mineral compositions: they reasonably assumed that chondrule density is imposed by their constituent minerals. Hughes (1977, 1978a) reported the aggregate bulk density of 955 disaggregated chondrules from the Bjurböle L/LL chondrite. He found a mean of 3.258 ± 0.008 g/cm$^3$. Hughes (1980) found that a subset of the chondrules used in his previous study appears to have a relationship between chondrule density and size. In an abstract,

Gooding et al. (1978) reported the mean density of 191 ordinary (H, L, LL) chondrite chondrules as 3.19 g/cm$^3$. Gooding (1983) provided more detail for individual chondrules and we refer the reader to that work for specifics. He found a range of 2.96 - 3.38 g/cm$^3$ for ordinary chondrite chondrules. A mean of 3.15 g/cm$^3$ can be calculated for the suites of 294 chondrules tabulated in the Gooding (1983) data. Teitler et al. (2010) measured the masses and derived the densities of hundreds of chondrules in H, L, L/LL (Bjurböle), and CV chondrites. They reported their data in terms of radius×density, an astrophysically-relevant (Stokes) parameter, so no density data are presented here.

## 5. Chondrule-like objects

### 5.1. Microchondrules, macrochondrules, and megachondrules

Occasionally, spherical (and predominately silicate) objects have apparent diameters that are significantly smaller or larger than established size distributions for a given chondrite chemical group. This is the case for OCs, but the recognition of microchondrules does extend to the CV$_{oxA}$ chondrite Allende (Fruland et al., 1978) and the CV$_{red}$ chondrite Vigarano (Rubin et al., 1982). Microchondrules, or chondrules that are orders of magnitude smaller in apparent diameter than the typical chondrules in a host chondrite, have been identified in several unequilibrated ordinary chondrites (Levi-Donati, 1970; Rubin et al., 1982; Krot, et al., 1997). They are typically found within rims of MgO-rich, FeO-poor (Type I) chondrules, but microchondrule-bearing lithic clasts apparently unassociated with chondrule rims have been noted (Rubin, 1989b). Krot et al. (1997) defined microchondrules as chondrules <40 μm in diameter, but others have used different definitions. Rubin et al. (1982) reported a lithic fragment containing chondrules ranging in diameter from 0.2-74μm in the Piancaldoli (LL3.4) chondrite. However, the material containing the microchondrules was later interpreted to be a chondrule rim rather than a lithic fragment (Krot and Rubin, 1996). Rubin et al. (1982) also noted similarly sized microchondrules in the Rio Negro L4 regolith breccia. They also performed a systematic search for microchondrules (defined as objects with 12-100 μm apparent diameters) in H, L, LL, CO, and CV chondrites yielding estimates for their abundances in each chondrite group. They inferred that microchondrules are most abundant in H and CO chondrites. Christophe Michel-Lévy (1987, 1988) described clasts in Mezö-Madaras (L3.7) that contained microchondrules between 3 and 100 μm diameter. She also noted that Krymka (LL3.2) contained microchondrules within fine grained regions and occasionally within sulfide grains. Rubin (1989b) also reported the occurrence of about thirty 3-31 μm (apparent diameter) microchondrules in a clast (that does not appear to be a chondrule rim) in the Krymka LL chondrite. More recently, Bigolski et al. (2014) and Dobrică and Brearley (2014) have reported on microchondrules in the LL3.0 Semarkona, ungrouped OC3.05 Northwest Africa (NWA) 5717, LL3.15 Bishunpur and MET 00526 (L3.05) chondrites.

The origin of microchondrules is under debate. It is not known whether they are the result of processes similar to those that formed chondrules more typical in size. Current possibilities for their formation include splattering or spalling of material from more typically sized chondrules, formation as protuberances from partially molten parent chondrules, or melting of FeO-rich dust in a process similar to that experienced by typical chondrules (Krot et al., 1997; Dobrică and Brearley, 2014; Bigolski et al., 2014).

Macrochondrules, spherical igneous-textured objects larger than typical host rock chondrules, have also been reported. Based on available literature, Weisberg et al. (1988a) concluded that chondrules in ordinary chondrites with apparent diameters of >4 mm are extremely rare. They defined a macrochondrule as a chondrule-like object with diameter >5 mm. Earlier, Binns (1967) found a large chondrule in the Parnallee LL chondrite. Weisberg et al. (1988a) reported the existence of seven macrochondrules in seven different unequilibrated and equilibrated OCs. Prinz et al. (1988) described a golf ball sized (2.5 cm diameter) igneous object in the Gunlock L3.2 chondrite. Ruzicka et al. (1998) reported the compositions of a number of "megachondrules", or exceptionally large chondrule-like objects in the unequilibrated Julesberg L3.6 chondrite. Weyrauch and Bischoff (2012) studied 74 chondrules with diameters of >3 mm. They found these objects in nearly all chemical classes of chondrites.

There is some consensus on the origins of macrochondrules. Most of the above have advocated that these formed by the same processes (but under different gas/dust ratios) as more average-sized chondrules (Binns, 1967; Weisberg et al., 1988a; Prinz et al. 1988) or by collisional coagulation of average-sized precursor chondrules (Weyrauch and Bischoff, 2012).

*5.2. Metal chondrules?*

Metal-rich chondrule-like objects (generally described as metal "spherules" or "nodules") have been reported in several groups of chondrites. These nodules are often associated with or contain sulfides. The origins of large (> ~2 mm) metal nodules and veins in equilibrated ordinary chondrites are clearly related to parent body or impact processing (Widom et al., 1986; Kong et al., 1998; Rubin, 1999; Friedrich et al., 2013). However, unequilibrated (and mildly shock processed) ordinary chondrites such as Bishunpur (LL3.1), Semarkona (LL3.0), and Watonga (LL3.1) also contain metallic Fe-Ni spherules (50-250 μm in apparent diameter) and these are likely of nebular origin (e.g., Rambaldi and Wasson, 1981). Wang et al. (2007) also favor a nebular origin for the 100-600μm diameter Fe-Ni metal spherules in the Ningqiang carbonaceous chondrite. Skinner and Leenhouts (1993) interpreted the chondrule-sized (740 μm mean apparent diameter) metal spherules as metal chondrules in the Acfer 059 CR chondrite. Weisberg et al. (2013) described metal nodules in EL3 chondrites as being 200-300 μm in apparent diameter (smaller than the chondrules in EL chondrites, see section 3.2.2.) and constituting about 10% of EL chondrite volumes. However, we point out that these metal nodules have a radius × density parameter comparable to EL chondrules, and this may favor a nebular origin. While Weisberg et al. (2013) favor a nebular origin for metallic nodules in EL chondrites, Van Niekerk and Keil (2011) proposed an impact origin.

It is unclear if the metal spherules or "chondrules" found in chondrites experienced the same astrophysical (heating) environment as their silicate cousins. However, some Fe-Ni metal (and sulfide) rich spherules – especially those in very unequilibrated or unaltered chondrites, are undoubtedly of nebular origin and investigations into their physical properties and origins are an area rich for new discoveries.

# 6. Other facets of chondrules
*6.1. Relationships between chondrule size, petrography, and composition.*

Chondrules possess a variety of petrographic textures. One convenient scheme of grouping the textures and compositions was proposed by Gooding and Keil (1981). Using this scheme, one can separate chondrules into porphyritic [porphyritic olivine (PO), porphyritic pyroxene (PP), porphyritic olivine-pyroxene (POP)] and non-porphyritic [barred olivine (BO), radial pyroxene (RP), cryptocrystalline (C), and granular olivine-pyroxene (GOP)] textures (e.g. Rubin, 1989a). Gooding et al. (1978) determined the percentage abundances of chondrule types from the L chondrite group and concluded that chondrule size and shape are not strictly correlated with chondrule type. Gooding and Keil (1981) and Gooding (1983) provided additional data on OC chondrule textural type by size and also found no correlation with textural type and size. They admitted that their conclusions may not have been statistically significant since they only studied 141 chondrules in total and as we have noted (above), their chondrules were not completely representative. Nagahara (1981) provided a plot of size distribution with respect to textural type in the ALH 77015 chondrite (L3.5, Fig. 2). She concluded that no relationship exists between texture and size, but pointed out that BO chondrules showed a hint of bias toward smaller sizes. Rubin and Keil (1984) found no statistically significant correlation of chondrule type with size in Inman (L/LL3.4) and ALH A77011 (L3.5). In an abstract, Goswami (1984) suggested that the frequency of non-porphyritic chondrules may increase at the lower end of the distribution of chondrules in OCs. Rubin and Grossman (1987) reported that in EH chondrites, RP chondrules are somewhat larger than C chondrules. They also found that non-porphyritic chondrules have a broader size-frequency distribution than porphyritic chondrules and that POP chondrules are significantly larger than PP chondrules. Rubin (1989a) found that porphyritic chondrules are statistically significantly larger than non-porphyritic chondrules in CO chondrites. Similar to the EH chondrites, Rubin (1989a) found that in CO chondrites, PO chondrules are larger than PP chondrules. In LL chondrites, Nelson and Rubin (2002) found the direct opposite: non-porphyritic chondrules are generally larger than porphyritic chondrules in the LL chondrites. In their examination of the textures of chondrule fragments, they found that porphyritic chondrules were more likely to be fragmented than non-porphyritic chondrules. They concluded that differences in size among chondrule textural types were primarily due to chondrule fragmentation events on the parent asteroid and not to chondrule formation processes in the solar nebula. This insight further complicates inferences about relationships between chondrule textural type and size-frequency distribution. In summary, there may be some relationship between chondrule size and textural type, but the data are sparse and is complicated by parent body processing.

Chondrules can also be grouped by their composition: Type I (low FeO and moderately volatile element poor) and Type II (high FeO content and less depleted in moderately volatile elements) (Hewins et al., 1996). In an abstract, Haack and Scott (1993) stated that type I chondrules are smaller than type II chondrules in the Roosevelt County 075 H3.2 chondrite, but more detailed size information was not given.

*6.2. Compound and cratered chondrules*

Compound chondrules are chondrules that are connected binary, ternary, or even quaternary (Friedrich, unpublished data) chondrules. Rubin (2010 and personal communication) categorizes compound chondrules into several types. Nested (or enveloping) compound chondrules are compound chondrules consisting of a spherical shell (the secondary chondrule) or shells (additional enveloping chondrules) around a primary chondrule. The grain size of the secondary or additional chondrules is akin to the primary chondrule. If the shell is finer grained, it may be acknowledged as an igneous rim rather than a compound chondrule (Rubin, 2010). Sibling compound chondrules are two or more chondrules of similar size that are attached to one another. Finally, adhering compound chondrules consist of a primary chondrule and one or more attached small chondrules at its surface.

Gooding and Keil (1981) were the first to report on the abundance of compound chondrules and what they referred to as cratered chondrules or chondrules with a bowl shaped depression. They observed a frequency of ≤3.5% for compound chondrules in ordinary chondrites and found that ≤1.5% of ordinary chondrite chondrules displayed cratering phenomena. They acknowledged that stereological sampling biases (they observed 2D petrographic thin sections) imply that these are lower limits for these phenomena. They concluded that <4% of all chondrules should be compound chondrules and that <10% of all chondrules should be cratered. Based on a study of 56 sets of compound chondrules in ordinary chondrites, Wasson et al. (1995) found a frequency of 2.4% for compound chondrules. They estimate that 58% of all OC compound chondrules are "siblings" (similar textures and compositions) and 42% are "independent" (they suggest different textures or compositions mean two individual chondrules formed from different batches of precursor material). Ciesla et al. (2004) used modeling and numerical arguments generated from 2D petrographic observations to suggest that 5% of chondrules in OCs are compound chondrules. Akaki and Nakamura (2005) found abundances of 1.6% and 0.4% compound chondrules in the Allende and Axtell CV chondrites respectively.

Compound chondrules (Wasson et al., 1995; Ciesla et al., 2004) and "cratered" chondrules may yield information about the chondrule formation environment or their early (while still plastic) history. Gooding and Keil (1981) suggested that plastic chondrules that collided with each other but subsequently separated are the origin of the cratered chondrules (a nebular origin). They inferred that the abundance of cratered chondrules could potentially be used to estimate chondrule number densities in a chondrule forming region. However, Grossman and Wasson (1985) suggested that the origins of "cratered" chondrules were actually locations on the chondrule where metal and sulfide droplets escaped while the chondrule was plastic. If this is the case, the abundances of cratered chondrules cannot be used to estimate number densities during chondrule formation. Similarly, inferring the spatial densities of chondrules from the occurrence of compound chondrules may be problematic because of the unknown timing of the addition of subsequent chondrules to the primary.

## 7. Evaluation
### 7.1. Recommended chondrule diameters

We have compiled literature data of reported chondrule sizes and size-frequency distributions across all types of chondrites and the acapulcoite primitive achondrites. Using our compilation in Table 7, we give some recommended values for typical chondrule diameters among all meteorite groups. We arrive at these values after considering all chondrule data and note that as discussed immediately above (Section 7.1), the majority of the means we report are not pure arithmetic means. Most values are based on log normal (phi) based data, which takes into account the asymmetric probability density function of chondrule size frequency. Arithmetic means will differ from those given here.

It is commonly accepted that average chondrule sizes increase from H to LL (H<L<LL) among the OCs. Our compilation reveals that, while this is true, the actual differences in mean diameters may not be as pronounced as previously accepted. Our recommended mean values for the OCs do increase H (~450 µm) – L (~500 µm) – LL (~550 µm) (Table 7). The medians for each group likely reside at 500 ±100 µm. Figure 7 illustrates reliable size frequency distributions for an H chondrite and four individual LL chondrite datasets. Since no statistically large chondrule size-frequency data for L chondrites are available (cf. Table 2), we use the L/LL chondrite Bjurböle as a proxy. However, we admit that it is unclear if the L/LL chondrites more closely resemble the L or LL chondrites. Figure 7 shows the increasing arithmetic mean is a result of an increasing (H-L-LL) positive (toward coarser chondrule sizes) skewness of the chondrule size-frequency distributions (Fig. 7) among the ordinary chondrite groups. This can also be seen in our typical maximum recommended chondrule diameters for the OCs (Table 7). A minimum chondrule diameter among the OCs probably exists – most studies would suggest ~100 µm as a common minimum cutoff, although smaller chondrules have been infrequently reported. We also give typical maximum chondrule diameters for the OCs: H (~1500 µm), L (~1900 µm), and LL (~ 2600 µm). These rule-of-thumb maximum diameters are based on the size frequency distributions and on the fact that 95% of chondrules in each chemical group will probably reside below that value (cf. Fig 7).

Among the EH and EL chondrites, EH chondrites have the best defined size-frequency distribution. The typical EH chondrite range is 50-1200 µm, with the typical max being defined as per above with the OCs. EH chondrites appear to have a mean chondrule diameter smaller than the OCs: our compilation suggests 230 µm is a reasonable value. EL chondrites have a mean diameter around ~500 µm.

The R chondrite mean diameter is based on limited data, which suggests ~400 µm as a reasonable mean. We are only able to give a range for the K chondrite mean diameter – current data suggest a mean residing between 500 and 1100 µm, but the true mean likely lies at the lower end of that range.

The CM mean chondrule apparent diameter is based on only one published study, which suggests a 270 µm mean diameter. The CO chondrite chondrule diameters have been well-documented as the 150 µm value shown in Table 7. The CV and CK chondrite mean diameter of 900 µm rests on relatively new data, but it has long been accepted that chondrules in CV chondrites are significantly larger than those in the OCs, for example. The CR chondrite mean value of 700 µm is also rather robust because of the variety of samples it is derived from. The CBb chondrite chondrules probably have a mean diameter of ~200µm, but it is not well constrained. It is also clear that the CH chondrites have chondrule-like objects that are an order of magnitude smaller than objects in other chondrite groups, but it is debated whether they (or the CB chondrules) are true chondrules or the result of another process, such as early impact processing.

### 7.2. Commentary
#### 7.2.1. Stereological correction of 2D petrographic data

As mentioned in Section 2 and elsewhere, since most chondrule dimensional data exist only in the form of 2D apparent diameters obtained from the study of petrographic thin sections, some investigators have utilized corrections for the bias between apparent 2D and true 3D diameters of chondrules. Dodd (1976) used the empirically-determined conversion curves of Friedman (1958). Hughes (1978a) applied a numerical correction, although his mathematical treatment assumed that chondrules were all of equal diameter, which they are clearly not. Eisenhour (1996) presented an improved means of correction, but his correction forces the chondrule size-frequency distribution into a Weibull distribution function. However, Teitler (2010) demonstrated that size distributions of disaggregated chondrules are not completely described by the Weibull distribution, questioning the validity of the Eisenhour (1996) correction. There is a need for a completely non-parametric 2D to 3D stereological corrections for chondrule sizes. Cuzzi (personal communication) has developed an applicable "unfolding" algorithm, but the complete presentation has yet to be described other than in abstract form (Christoffersen et al., 2012).

*7.2.2. Suggestions and future prospects*

As discussed above, the vast majority of chondrule dimensional data exist in the form of 2D apparent diameters obtained from the study of petrographic thin sections. No tested, reliable stereological correction is at hand, so the true 3D dimensions of chondrules remain obscured. In cases when disaggregation were performed and separated chondrules were measured, there is warranted concern about potential biases in the datasets either inadvertently (see Teitler et al., 2010) or because disaggregation studies were often incidental, with the true goal being a compositional study for which larger chondrules were selected for ease of handling.

Researchers are encouraged to more specifically state the statistical assumptions used for presenting their data. There are three general approaches used: calculating statistics arithmetically (assuming a normal distribution with metric size values), logarithmically (assuming a log-normal distribution and using phi size values), and geometrically (assuming a log-normal distribution with metric size values). Often, readers can only assume that an arithmetic mean was calculated in cases where symmetrical standard deviations are reported. Rubin (1989a) has generally used and stated the assumptions behind his presentations of data: phi (φ) units [-log2 diameter (mm)] are used as they approximate the apparent log normal distribution of chondrule diameters, and means and asymmetric standard deviations are then given. As noted above, however, the true statistical probability distribution function that describes the size-frequency distribution of chondrules remains mysterious. An accurate assessment of the true statistical distribution that describes chondrule sizes would undoubtedly benefit from consistent reporting of undigested (raw) chondrule dimensions for future researchers to use for hypothesis testing.

Improvement of our knowledge of chondrule size distributions without stereological correction is recommended. Today, this can be accomplished with 3D methods such as x-ray microtomography (μCT) (e.g., Ebel and Rivers, 2007). The use of μCT has the potential to revolutionize the measurement of chondrule size frequency distributions since disaggregation and the associated uncertainties such as loss of material or adherence of matrix during disaggregation are minimized. However, significant challenges remain in the automated digital segmentation and separation of chondrules within μCT volumes because of the extremely heterogeneous textures and composition (densities) of chondrules, even in a single stone. At this time, human intervention in segmentation of chondrules in μCT volumes remains necessary for accurate and precise chondrule dimension determination (Friedrich, 2014).

## 8. Conclusions

We have compiled available chondrule dimensional data from the literature for all primitive meteorite groups. Based on our compiled data, we have provided recommended values for the mean diameters of chondrules in each of the chondrite groups. Chondrules have approximately log-normal size distributions, but their authentic size-frequency probability density function is unknown. We find that the OCs have increasing mean chondrule diameters: H (~450 μm) – L (~500 μm) – LL (~550 μm). These robust recommended values are less extreme than previously thought. Other chondrite groups (EL, R, K) display mean diameters near 500 μm, but EH chondrite chondrules are about ~230 μm in mean diameter. Carbonaceous chondrites represent a chemically-diverse collection of primitive parent bodies and their chondrules are likewise diverse in average diameter, ranging from ~150 to 900 μm (Table 6, Fig. 6). It is generally accepted that the CH chondrite chondrules did not form by the same processes as other chondrules. This conclusion is partially based on their smaller (~20 μm) mean diameter.

True individual chondrule density measurements can only be accomplished by the disaggregation of chondrites, which explains why so few such data exist. However, the few studies that report chondrule density data suggest that OC chondrite chondrules have densities between 3.15 and 3.26 $g/cm^3$, as may be expected from their mineral and glass compositions.

The existence of metal and sulfide chondrules is controversial, but further investigation of the idea is warranted and may yield insights into the astrophysical formation environment of their silicate counterparts and comparisons in OCs. Similarly, the frequency of observation of other textural aspects of chondrules, such as compound chondrules, "cratered" chondrules, and the sizes of micro- and macrochondrules may also provide additional constraints on chondrule formation processes. There seem to be few systematic relationships between chondrule petrographic texture, composition, and size across all chondrite groups. Some statistically significant correlation between textural type and chondrule size within the LL, EH, and CO chondrites appear to emerge from the data when each is considered individually. However, no particular textural type of chondrule seems consistently smaller or larger across multiple chondrite groups.

A majority of chondrites have chondrule mean diameters near 500 μm in diameter. Chondrules are often referred to as "mm-sized" silicate spherules; however, a better description may be "half-mm-sized" spherules. While much has been learned with respect to chondrule size distributions, there is still significant knowledge remaining to be acquired. The prospect of true 3D data with the use of μCT holds several advantages over traditional 2D petrographically-collected data since it does not require stereological correction.

## Acknowledgments

Discussions with Drs. J. Grossman, A. Rubin, J. Cuzzi, J. Paque, J. Simon, and K. Kuebler were exceptionally helpful. J. Bigolski provided assistance with navigating the microchondrule literature. This Invited Review was solicited and handled by Associate Editor Klaus Keil. The work of MKW is supported by NASA Cosmochemistry grant NNX12AI06G. DSE acknowledges funding from NASA grant NNX10AI42G. Timely and helpful reviews by Drs. A. Rubin and J. Paque were especially welcome.

Table 1. Summary of published H and H/L chondrite chondrule diameter data.

| chondrite | pet. type | reference | n[a] | mean (µm) | median (µm) | range (µm) | method[b] | 2D→3D correction[c] | notes |
|---|---|---|---|---|---|---|---|---|---|
| Tieschitz | H/L3.6 | Dodd 1976 | 130 | | 420 | | PTS | Y[d] | H/L , includes chondrule fragments |
| Sharps | 3.4 | Dodd 1976 | 254 | | 290 | | PTS | Y[d] | includes chondrule fragments |
| Sharps | 3.4 | Dodd 1976 | 233 | | 280 | | PTS | Y[d] | includes chondrule fragments |
| Bremervörde | H/L3.9 | Dodd 1976 | 114 | | 320 | | PTS | Y[d] | H/L, includes chondrule fragments |
| Sindhri | 5 | Dodd 1976 | 160 | | 340 | | PTS | Y[d] | includes chondrule fragments |
| Prairie Dog Creek | 3.7 | Dodd 1976 | 154 | | 280 | | PTS | Y[d] | includes chondrule fragments |
| Clovis (no. 1) | 3.6 | Dodd 1976 | 230 | | 420 | | PTS | Y[d] | includes chondrule fragments |
| Selma | 4 | Dodd 1976 | 341 | | 300 | | PTS | Y[d] | includes chondrule fragments |
| Allegan | 5 | Martin & Mills 1978 | 1256 | 570 | 600 | 150-2750 | D | | |
| Bremervörde | H/L 3.9 | King & King 1979 | 56 | | 510 | | PTS | N | |
| Clovis (no. 1) | 3.6 | King & King 1979 | 153 | | 330 | | PTS | N | |
| Dimmitt | 3.7 | King & King 1979 | 32 | | 280 | | PTS | N | |
| Prairie Dog Creek | 3.7 | King & King 1979 | 104 | | 370 | | PTS | N | |
| Tieschitz | H/L 3.6 | King & King 1979 | 46 | | 530 | | PTS | N | |
| Weston | 4 | King & King 1979 | 69 | | 340 | | PTS | N | |
| various H chondrites | 3 | Lux et al. 1981 | 87 | 420 | | | D | | weighted mean diameter (see text) |
| Tieschitz | H/L 3.6 | Gooding 1983 | 26 | 1090 +410/ -300e | | | D | | H/L, compositional study, size bias evident and noted by author |
| Dhajala | 3.8 | Gooding 1983 | 14 | 1000 +360/ -270e | | | D | | compositional study, size bias evident and noted by author |
| Weston | 4 | Gooding 1983 | 16 | 860 +350/ -250e | | | D | | compositional study, size bias evident and noted by author |
| Ochansk | 4 | Gooding, 1983 | 9 | 930 +490/ -320e | | | D | | compositional study, size bias evident and noted by author |
| - | | Grossman et al. 1988a | | 300 | | | - | - | estimated mean from literature compilation |
| Hammond Downs | 4 | Kuebler et al. 1999 | 261 | 460±12 | | | PTS | Y[f] | corrected mean value shown , uncorrected values not available. |
| GRO 95524 | 5 | Teitler et al. 2010 | 300 | 514±220 | 470 | 150-1326 | D | | |

[a] n = number of chondrules considered in the study, blank if number of chondrules was not reported
[b] PTS = petrographic thin section, D = disaggregation;
[c] In the case of petrographic thin section measurements, indicates if a stereological correction was applied to the data. Not applicable to disaggregation studies.
[d] Friedman (1958) empirical correction applied
[e] geometric mean
[f] Eisenhour (1996) correction applied

Table 2. Summary of published L chondrite chondrule diameter data.

| chondrite | pet. type | reference | n[a] | mean (μm) | median (μm) | range (μm) | method[b] | 2D→3D correction[c] | notes |
|---|---|---|---|---|---|---|---|---|---|
| Elenovka | 5 | Stakheav et al. 1973 | 637 | | | | D | | includes chondrule fragments, see Fig. 2 and text. |
| Nikolskoe | 4 | Stakheav et al. 1973 | 1090 | | | | D | | includes chondrule fragments, see Fig. 2 and text. |
| Saratov | 4 | Stakheav et al. 1973 | 3714 | | | | D | | includes chondrule fragments, see Fig. 2 and text. |
| Hallingeberg | 3.4 | Dodd 1976 | 242 | | 420 | | PTS | Y[d] | includes chondrule fragments |
| Mezö-Madaras | 3.7 | Dodd 1976 | 687 | | 420 | | PTS | Y[d] | includes chondrule fragments |
| Khohar | 3.6 | Dodd 1976 | 367 | | 400 | | PTS | Y[d] | includes chondrule fragments |
| Carraweena | 3.9 | Dodd 1976 | 354 | | 500 | | PTS | Y[d] | includes chondrule fragments |
| Ioka | 3.5 | Dodd 1976 | 150 | | 490 | | PTS | Y[d] | includes chondrule fragments |
| Barratta | 4 | Dodd 1976 | 279 | | 550 | | PTS | Y[d] | includes chondrule fragments |
| Khohar | 3.6 | King & King 1979 | 132 | | 620 | | PTS | N | |
| Mezö-Madaras | 3.7 | King & King 1979 | 58 | | 490 | | PTS | N | |
| Y-74191 | 3.7 | Ikeda & Takeda 1979 | 119 | | | 500-700 | PTS | N | range of means, >200 μm chondrules only |
| ALH A77015 | 3.5 | Nagahara 1981 | 108 | ~800 | | | PTS | N | |
| Hallingeberg | 3.4 | Gooding 1983 | 22 | 890 +310/-230e | | | PTS | N | size bias evident and noted by author |
| Saratov | 4 | Gooding 1983 | 13 | 1160 +510/-360e | | | PTS | N | size bias evident and noted by author |
| Saratov | 4 | Gooding 1983 | 20 | 1080 +530/-360e | | | PTS | N | size bias evident and noted by author |
| Tennasilm | 4 | Gooding 1983 | 6 | 900 +400/-270e | | | PTS | N | size bias evident and noted by author |
| Tennasilm | 4 | Gooding 1983 | 12 | 920 +470/-310e | | | PTS | N | size bias evident and noted by author |
| ALH A77011 | 3.5 | Rubin & Keil 1984 | 163 | 680±625 | | 90-5080 | PTS | N | Barred Olivine (BO), abbreviated size range 130-1900 μm |
| ALH A77011 | 3.5 | Rubin & Keil 1984 | 70 | 622±453 | | 73-1780 | PTS | N | Radial Pyroxene (RP) and Cryptocrystalline (CC), abbreviated size range 77-1770 μm |
| ALH A77011 | 3.5 | Rubin & Grossman 1987 | | 476 +554/-255 | | | | | |
| - | - | Grossman et al. 1988a | | 600-800 | | | - | - | estimated mean from literature compilation |
| ALH 85033 | 4 | Paque & Cuzzi 1997; Cuzzi et al. 1999 | 235 | 720 | | | D | | |
| ALH 85033 | 4 | Tietler et al. 2010 | 235 | 462±260 | 384 | 174-1898 | D | | chondrules massed, same chondrules as Paque & Cuzzi (1997) and Cuzzi et al. (1999) and Cuzzi et al. (2001) |
| NWA 869 | 3-6 (<3.5)[f] | Metzler 2012 | 67 | 520 | | 100-1300 | PTS | N | |

[a] n = number of chondrules considered in the study, blank if number of chondrules was not reported

[b] PTS = petrographic thin section, D = disaggregation

[c] In the case of petrographic thin section based measurements, indicates if a stereological correction was applied to the data. Not applicable to disaggregation studies.

[d] Friedman (1958) empirical correction applied

[e] geometric mean;   [f] studied a clast described in Metzler et al. (2011)

Table 3. Summary of published Bjurböle (L/LL 4) and Inman (L/LL 3.4) chondrite chondrule diameter data.

| chondrite | pet. type | reference | n[a] | mean (µm) | median (µm) | range (µm) | method[b] | 2D→3D correction[c] | notes |
|---|---|---|---|---|---|---|---|---|---|
| Bjurböle | 4 | Stakheav et al. 1973 | 997 | | | | D | | includes chondrule fragments, see Fig. 3 and text. |
| Bjurböle | 4 | Dodd 1976 | 272 | | 260 | | PTS | Y[d] | includes chondrule fragments |
| Bjurböle | 4 | Martin & Mills 1976 | 97 | 1180 ±1110 | 1120 | 400-2200 | D | | |
| Bjurböle | 4 | Hughes 1977, 1978a | 61 | 817 | 843 | 200- ~1600 | PTS | Y[e] | uncorrected mean 653 µm, uncorrected median 678 ± 5 µm |
| Bjurböle | 4 | Hughes 1977, 1978a | 955 | 750 | 688±3 | 250-3670 | D | | |
| Bjurböle | 4 | Hughes 1980 | 176 | | | 300-3200 | D | | density study where only highly spherical chondrules included |
| Inman | 3.4 | King & King 1979 | 118 | | 600 | | PTS | N | |
| Inman | 3.4 | Rubin and Keil 1984 | 173 | 1038 ±937 | | 140-5973 | PTS | N | barred olivine (BO) chondrules, abbreviated size range (second smallest to second largest) 170-5600 µm |
| Inman | 3.4 | Rubin and Keil 1984 | 201 | 852 ±598 | | 48-4278 | PTS | N | radial pyroxene (RP) and cryptocrystalline (CC) chondrules, abbreviated size range 90-3667 µm |
| Inman | 3.4 | Rubin & Grossman 1987 | | 688 +664/ -338 | | | | | |
| Bjurböle | 4 | Kuebler et al. 1999 | 210 | 590 ±250 | 590 | | PTS | Y[f] | uncorrected mean 573 ± 320 µm, uncorrected median 522 µm |
| Bjurböle | 4 | Teitler et al 2010 | 150 | 514±220 | 470 | 150-1326 | D | | possible sampling bias noted by authors |

[a] n = number of chondrules considered in the study, blank if number of chondrules was not reported; [b] PTS = petrographic thin section, D = disaggregation; [c] In the case of petrographic thin section measurements, indicates if a stereological correction was applied to the data. Not applicable to disaggregation studies.
[e] method for correction outlined in Hughes (1978a)
[d] Friedman (1958) empirical correction applied
[f] Eisenhour (1996) correction applied

Table 4. Summary of published LL chondrite chondrule diameter data.

| chondrite | pet. type | reference | n[a] | mean (µm) | median (µm) | range (µm) | method[b] | 2D→3D correction[c] | notes |
|---|---|---|---|---|---|---|---|---|---|
| Bishunpur | 3.15 | Dodd 1976 | 153 | | 340 | | PTS | Y[d] | includes chondrule fragments |
| Bishunpur | 3.15 | Dodd 1976 | 118 | | 400 | | PTS | Y[d] | includes chondrule fragments |
| Chainpur | 3.4 | Dodd 1976 | 96 | | 470 | | PTS | Y[d] | includes chondrule fragments |
| Hamlet | 4 | Dodd 1976 | 118 | | 510 | | PTS | Y[d] | includes chondrule fragments |
| Krymka | 3.2 | Dodd 1976 | 294 | | 530 | | PTS | Y[d] | includes chondrule fragments |
| Ngawi | 3.6 | Dodd 1976 | 157 | | 370 | | PTS | Y[d] | includes chondrule fragments |
| Parnallee | 3.6 | Dodd 1976 | 420 | | 510 | | PTS | Y[d] | includes chondrule fragments |
| Chainpur | 3.4 | Martin & Mills 1976 | 245 | 1090 | 1020 | | D | | |
| Chainpur | 3.4 | Hughes 1978a | 84 | 893 | 817 | | PTS | Y[e] | uncorrected mean 714 µm, uncorrected median 657 ± 5 µm |
| various | 3 and 4 | Gooding et al. 1978; Gooding & Keil 1981 | 70 | 1280 | | | | N | |
| Parnallee | 3.6 | King & King 1979 | 45 | | 366 | | PTS | N | "fluid drop" (round) chondrules only |
| Bishunpur | 3.15 | King & King 1979 | 28 | | 637 | | PTS | N | "fluid drop" (round) chondrules only |
| Piancaldoli | 3.6 | Rubin et al 1982 | | | | 140-1700 | PTS | N | |
| Piancaldoli (clast) | 3.6 | Rubin et al 1982 | 81 | 18 | | 3-64 | PTS | N | Data shown for optically identified microchondrules. Additional chondrules as small as 0.25µm were identified with scanning electron microscopy. |
| Semarkona | 3.00 | Gooding 1983 | 15 | 1390 +890/ -540[f] | | | PTS | | |
| | 3.00 | Gooding 1983 | 17 | 1280 +890/ -530[f] | | | PTS | | |
| Chainpur | 3.4 | Gooding 1983 | 14 | 1590 +240/ -210f | | | PTS | | |
| | 3.4 | Gooding 1983 | 20 | 1390 +380/ -300f | | | PTS | | |
| Hamlet | 4 | Gooding 1983 | 8 | 940 +290/ -220f | | | PTS | | |
| | 4 | Gooding 1983 | 10 | 940 +260/ -200f | | | PTS | | |
| Soko-Banja | 4 | Gooding 1983 | 7 | 1530 +250/ -210f | | | PTS | | |
| | 4 | Gooding 1983 | 15 | 1320 +460/ -340f | | | PTS | | |
| - | - | Grossman et al. 1988a | | 900 | | | | | estimated mean from literature compilation |
| Semarkona | 3.00 | Huang et al. 1996 | 190 | 752±338 | 691 | 244-2264 | PTS | N | |
| Krymka | 3.2 | Huang et al. 1996 | 96 | 698±284 | 646 | 270-1481 | PTS | N | |
| Kelly | 4 | Kuebler et al. 1999 | 222 | 660±18 | | | PTS | Y[g] | corrected data given and shown in Fig. 4 |
| Semarkona | 3.2 | Nelson & Rubin 1999 | 236 | 560 +430/ -240 | | 105 (min.) | | | |

| | | | | | | | | |
|---|---|---|---|---|---|---|---|---|
| Semarkona | 3.2 | Nelson & Rubin 2002 | 380 | 610 +1060/ -350 | 110-2470 | PTS | N | |
| Bishunpur | 3.15 | Nelson & Rubin 2002 | 86 | 590 +940/ -370 | 190-2360 | PTS | N | |
| Krymka | 3.2 | Nelson & Rubin 2002 | 91 | 520 +910/ -300 | 120-3110 | PTS | N | |
| Piancaldoli | 3.6 | Nelson & Rubin 2002 | 87 | 600 +910/ -400 | 170-1630 | PTS | N | |
| LEW 88175 | 3.4 | Nelson & Rubin 2002 | 75 | 440 +740/ -260 | 130-1590 | PTS | N | |
| LL chondrites | LL | Nelson & Rubin 2002 | 719 | 570 +980/ -340 | | PTS | N | |
| NWA 5206 | 3.05 | Metzler 2012 | 49 | 670 | 200-1500 | PTS | N | unusual clast |
| NWA 1756 | 3.10 | Metzler 2012 | 40 | 720 | 300-1700 | PTS | N | unusual clast |
| Krymka | 3.2 | Metzler 2012 | 35 | 600 | 200-1100 | PTS | N | unusual clast |
| NWA 5205 | 3.2 | Metzler 2012 | 99 | 1380 | 400-2800 | PTS | N | unusual clast |
| NWA 5205 | 3.2 | Metzler 2012 | 58 | 900 | 300-1500 | PTS | N | unusual clast |
| NWA 5205 | 3.2 | Metzler 2012 | 47 | 980 | 400-1800 | PTS | N | unusual clast |
| NWA 4572 | 3 | Metzler 2012 | 52 | 820 | 200-2400 | PTS | N | unusual clast |

[a] n = number of chondrules considered in the study, blank if number of chondrules was not reported
[b] PTS = petrographic thin section, D = disaggregation
[c] In the case of petrographic thin section based measurements, indicates if a stereological correction was applied to the data.  Not applicable to disaggregation studies.
[d] Friedman (1958) empirical correction applied
[e] method for correction outlined in Hughes (1978a)
[f] geometric mean and standard deviation
[g] Eisenhour (1996) correction applied

Table 5. Summary of published enstatite and other non-carbonaceous chondrite chondrule diameter data.

| chondrite | chem. / pet. type | reference | n[a] | mean (μm) | median (μm) | range (μm) | method[b] | 2D→3D correction[c] | notes |
|---|---|---|---|---|---|---|---|---|---|
| Qingzhen | EH3 | Rubin & Grossman 1987 | 63 | | | | D | N | bias noted by investigators, see Fig. E for histogram |
| ALH A77156, Kota-Kota, Qingzhen | EH | Rubin & Grossman 1987 | | 213 +277/ -120 | | | PTS | N | best value for all chondrule data, see Fig. E for histogram |
| various | EH | Grossman et al. 1988a | | 200 | | | PTS | | |
| ALH 84170, PCA 91085, PCA 91238 | EH | Schneider et al. 1998, 2002 | 135 | 278±229 | | 45-1313 | PTS | N | |
| various | EH | Rubin 2000 | | 220 | | | | | |
| ALH 85119, MAC 88180, PCA 91020 | EL | Schneider et al. 1998,2002 | 199 | 476±357 | | 85-2125 | PTS | N | |
| various | EL | Rubin 2000 | | 550 | | | PTS | N | |
| Acfer 217 | R3.8-5 | Kallemeyn et al. 1996 | 59 | 410 +220/ -140 | | | PTS | N | |
| ALH 85151 | R3.6 | Kallemeyn et al. 1996; also see Rubin and Kallemeyn 1989 | 38 | 410 +390/ -200 | | | PTS | N | |
| Carlisle Lakes | R3.8 | Kallemeyn et al. 1996;also see Rubin and Kallemeyn 1989 | 55 | 460 +330/ -190 | | | PTS | N | |
| PCA 91002 | R3.8-6 | Kallemeyn et al. 1996 | 42 | 310 +220/ -130 | | | PTS | N | |
| Rumuruti | R3.8-6 | Kallemeyn et al. 1996 | 28 | 360 +250/ -150 | | | PTS | N | |
| Y-75302 | R3.8 | Kallemeyn et al. 1996 | 14 | 340 +170/ -110 | | | PTS | N | |
| Y-793575 | R3.8 | Kallemeyn et al. 1996 | 23 | 350 +200/ -130 | | | PTS | N | |
| Y-82002 | R3.9 | Kallemeyn et al. 1996 | 4 | 370 +320/ -170 | | | PTS | N | |
| Y-82002 | R3.9 | Nakamura et al. 1993 | 22 | | | 200-500 | PTS | N | also one 3000 μm diameter chondrule |
| Kakangari | K3 | Weisberg et al. 1996 | | 690 | | | PTS | N | |
| LEW 87232 | K | Weisberg et al. 1996 | | 480 | | | PTS | N | |
| Lea County 002 | K3 | Weisberg et al. 1996 | | 1100 | | | PTS | N | also one 5300 μm diameter chondrule |

[a] n = number of chondrules considered in the study, blank if number of chondrules was not reported; [b] PTS = petrographic thin section, D = disaggregation; [c] In the case of petrographic thin section measurements, indicates if a stereological correction was applied to the data. Not applicable to disaggregation studies.

Table 6. Summary of carbonaceous chondrite chondrule diameter data.

| chondrite | chem. / pet. type | reference | n[a] | mean (µm) | median (µm) | range (µm) | method[b] | 2D→3D correction[c] | notes |
|---|---|---|---|---|---|---|---|---|---|
| Murray | CM2 | Rubin & Wasson 1986 | 100 | 270 ±240 | | | PTS | N | |
| Ornans | CO3.4 | King & King, see Rubin and Wasson 1988 | | 196 +122/ -75 | | | PTS | N | |
| various | CO | Rubin 1989a | 2834 | 148 +132/ -70 | | | PTS | N | |
| ALH A77307 | CO3.0 | May et al., 1999 | | 259±161 | | | PTS | N | |
| Lancé | CO3.5 | May et al., 1999 | | 297±156 | | | PTS | N | |
| Warrenton | CO3.7 | May et al., 1999 | | 289±126 | | | PTS | N | |
| Acfer 374 | CO3 | Moggi-Cecchi et al. 2006 | | 110 | | | PTS | N | |
| - | CV | McSween, 1977 | | | | 500-2000 | | | "acknowledged" (McSween, 1977) range |
| - | CV | Grossman et al. 1988a | | 1000 | | | | | estimated mean CV chondrites |
| Allende | CV3$_{oxA}$ | Paque and Cuzzi, 1997 | | 850 | | | D | | |
| ALH 84028 | CV3 | Paque and Cuzzi, 1997 | | 970 | | | D | | |
| Vigarano | CV3$_{red}$ | May et al., 1999 | | 680±416 | | | PTS | N | |
| Efremovka | CV3$_{red}$ | May et al., 1999 | | 655±545 | | | PTS | N | |
| Mokoia | CV3$_{oxB}$ | May et al., 1999 | | 683±535 | | | PTS | N | |
| Leoville | CV3$_{red}$ | May et al., 1999 | | 823±649 | | | PTS | N | |
| ALH 84028 | CV3 | Teitler et al. 2010 | 194 | 932±488 | 788 | 286-3660 | D | N | |
| Allende | CV3 | Teitler et al. 2010 | 287 | 912±644 | 780 | 266-9100 | D | N | |
| Allende | CV3 | Teitler et al. 2010 | 126 | 918±744 | 632 | 274-3960 | D | N | |
| various | CK | Kallemeyn et al. 1991 | | 500-750 | | | PTS | N | range of probable mean |
| HaH 337 | CK4 | Moggi-Cecchi et al. 2006 | | 700 | | | PTS | N | |
| various | CK | Greenwood et al., 2010 | | 700-879 | | | PTS | N | range of probable mean |
| NWA 1559 | CK3 | Rubin 2010 | 36 | 890±480 | 870 | 240-7520 | PTS | N | 2 anomalously large chondrules excluded from mean (3150, 7520µm) |
| Watson 002 | CK3-an | Geiger et al. 1993 | 43 | 870±380 | | 160-2100 | PTS | N | |
| DaG 431 | CK3-an | Zipfel et al. 2000 | | | | ~200->1000 | PTS | N | |
| NWA 1559 | CK3-an | Brandstätter et al., 2003 | | | | <500-2000 | PTS | N | |
| NWA 1560 | CK4/5 | Bukovanská et al., 2003 | | | | <500-2000 | PTS | N | |
| NWA 1563 | CK4 | Bukovanská et al., 2003 | | | | <500-2000 | PTS | N | |
| Kobe | CK | Tomeoka et al. 2005 | | 750 | | 500-2000 | PTS | N | |
| various | CR | Bischoff et al. 1992 | 188 | 1000±600 | | | PTS | N | |
| Acfer 059 | CR | Skinner & Leenhouts 1993 | 64 | 740±320 | | | | | metal rich chondrules |
| Acfer 059 | CR | Skinner & Leenhouts 1993 | 412 | 1440±580 | | | | | silicate rich chondrules |
| Renazzo | CR | Kallemeyn et al. 1994 | 50 | 690 +840/ -380 | | 84-2240 | PTS | N | |
| EET 87770 | CR | Kallemeyn et al. 1994 | 35 | 770 +740/ -380 | | 260-4400 | PTS | N | |

| | | | n[a] | | | PTS[b] | N[c] | |
|---|---|---|---|---|---|---|---|---|
| PCA 91082 | CR | Kallemeyn et al. 1994 | 34 | 770 +700/ -370 | 80-1890 | PTS | N | |
| Acfer 187 | CR | Kallemeyn et al. 1994 | 36 | 590 +770/ -330 | 155-2920 | PTS | N | |
| MAC 87320 | CR | Kallemeyn et al. 1994 | 41 | 490 +790/ -300 | 57-2460 | PTS | N | brecciation may have altered reported chondrule diameters |
| various | CR | Rubin 2000 | | 700 | | | | best mean |
| ALH 85085 | CH | Scott 1988 | | 20 | <4 -200 | PTS | N | |
| ALH 85085 | CH | Grossman et al. 1988b | | 20 +19/ -10 | ~1000 max. | PTS | N | |
| Acfer 182 (and pairs) | CH | Bischoff et al. 1993b | 202 | 90 ± 60 | 1100 max. | PTS | N | |
| Acfer 366 | CH | Moggi-Cecchi et al. 2006 | 170 | 110 | 35-450 | PTS | N | |
| Ischeyevo | CH-CB (breccia) | Ivanova et al. 2008 | | 100 | 20-400 | PTS | N | metal-rich lithology (CH-like) |
| Ischeyevo | CH-CB (breccia) | Ivanova et al. 2008 | | 400 | 100-1000 | PTS | N | metal-poor lithology (CB-like) |
| various | $CB_a$ | Weisberg et al. 2001 | | | ≤10000 (1 cm) | PTS | N | |
| various | $CB_b$ | Weisberg et al. 2001 | | 200 | ≤1000 | PTS | N | |
| various | CB | Weisberg et al. 2006 | | | 20-1000 | PTS | N | |

[a] n = number of chondrules considered in the study, blank if number of chondrules was not reported
[b] PTS = petrographic thin section,  D = disaggregation
[c] In the case of petrographic thin section measurements, indicates if a stereological correction was applied to the data.  Not applicable to disaggregation studies.

Table 7. Some recommended values of chondrule diameters for different chondrite and primitive achondrite groups.

| type | approximate mean (μm) | typical observed range (μm) | typical [a] max(μm) | sources |
|---|---|---|---|---|
| H | 450 | 100-1500 | ~1500 | King & King (1979), Kuebler et al. (1999), Teitler et al. (2010) |
| L | 500 | 100-1900 | ~1900 | Rubin & Grossman (1987), Teitler et al. (2010) |
| LL | 550 | 100-2600 | ~2600 | Nelson & Rubin (2002) |
| EH | 230 | 50-1200 | ~1200 | Rubin & Grossman (1987), Schneider et al. (2002) |
| EL | 500 | | | Rubin (2000), Schneider et al. (2002) |
| R | 400 | | | Kallemeyn et al. (1996) |
| K | 500-1100? | | | Weisberg et al. (1996) |
| CM | 270 | | | Rubin & Wasson (1986) |
| CO | 150 | | | Rubin (1989a) |
| CV, CK | 900 | | | Rubin (2010), Teitler et al. (2010) |
| CR | 700 | | | Kallemeyn et al. (1996) |
| CH | 20 | | | Grossman et al. (1998b) |
| CB | 200 | | | Weisberg et al. (2001) |
| acapulcoites | - | 400-700 | | McCoy et al. (1996), Rubin (2007) |

[a] maximum diameter of a chondrule with ≥5% abundance

Fig. 1. H chondrite chondrule diameters and size-frequency distributions. Abscissa has the same scale as Figs. 2-5 for comparison. The H chondrites probably have a mean chondrule diameter close to ~450 µm. See Table 1 for numerical data and methodology notes.

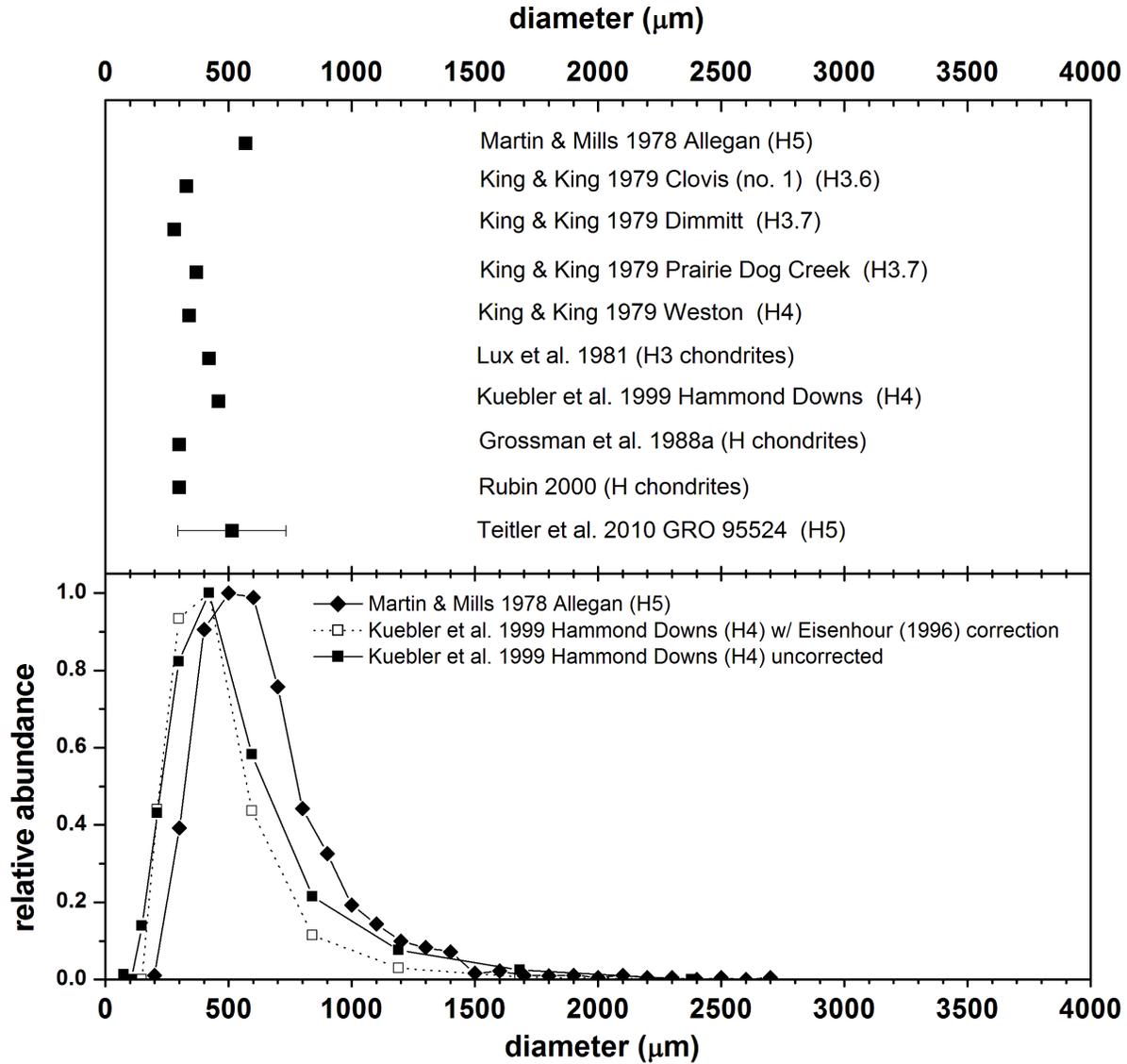

Fig. 2. L chondrite chondrule diameters and size-frequency distributions. Abscissa has the same scale as Figs. 1,3-5. L chondrite chondrules display a typical mean diameter of ~500 µm. See Table 2 for numerical data and commentary.

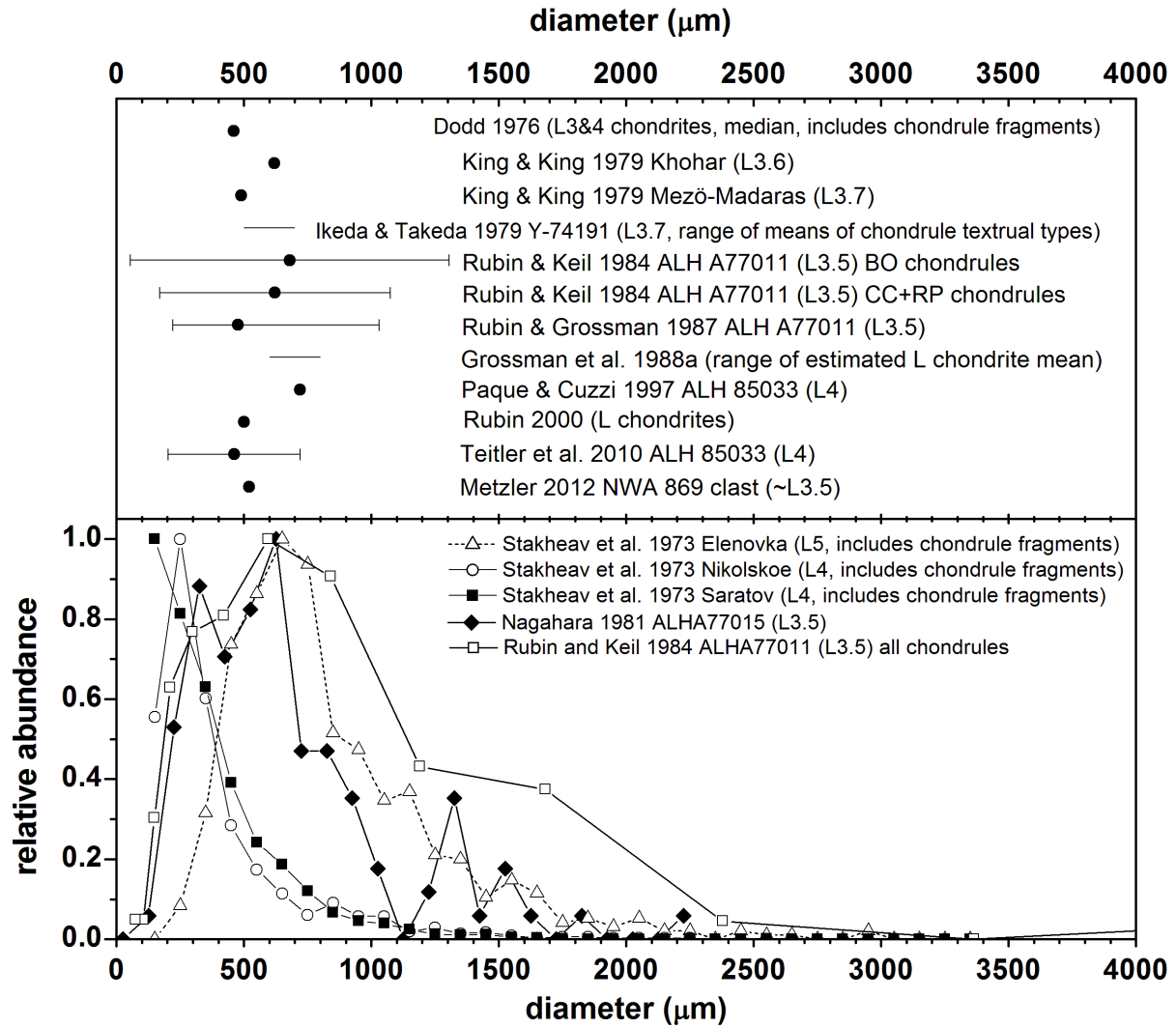

Fig. 3. Chondrule diameters and size-frequency distributions reported for the Bjurbole (L/LL) chondrite and (where noted) for the Inman (L/LL) chondrite. Abscissa has the same scale as Figs. 1-2,4-5 for comparison. See Table 3 for data and commentary.

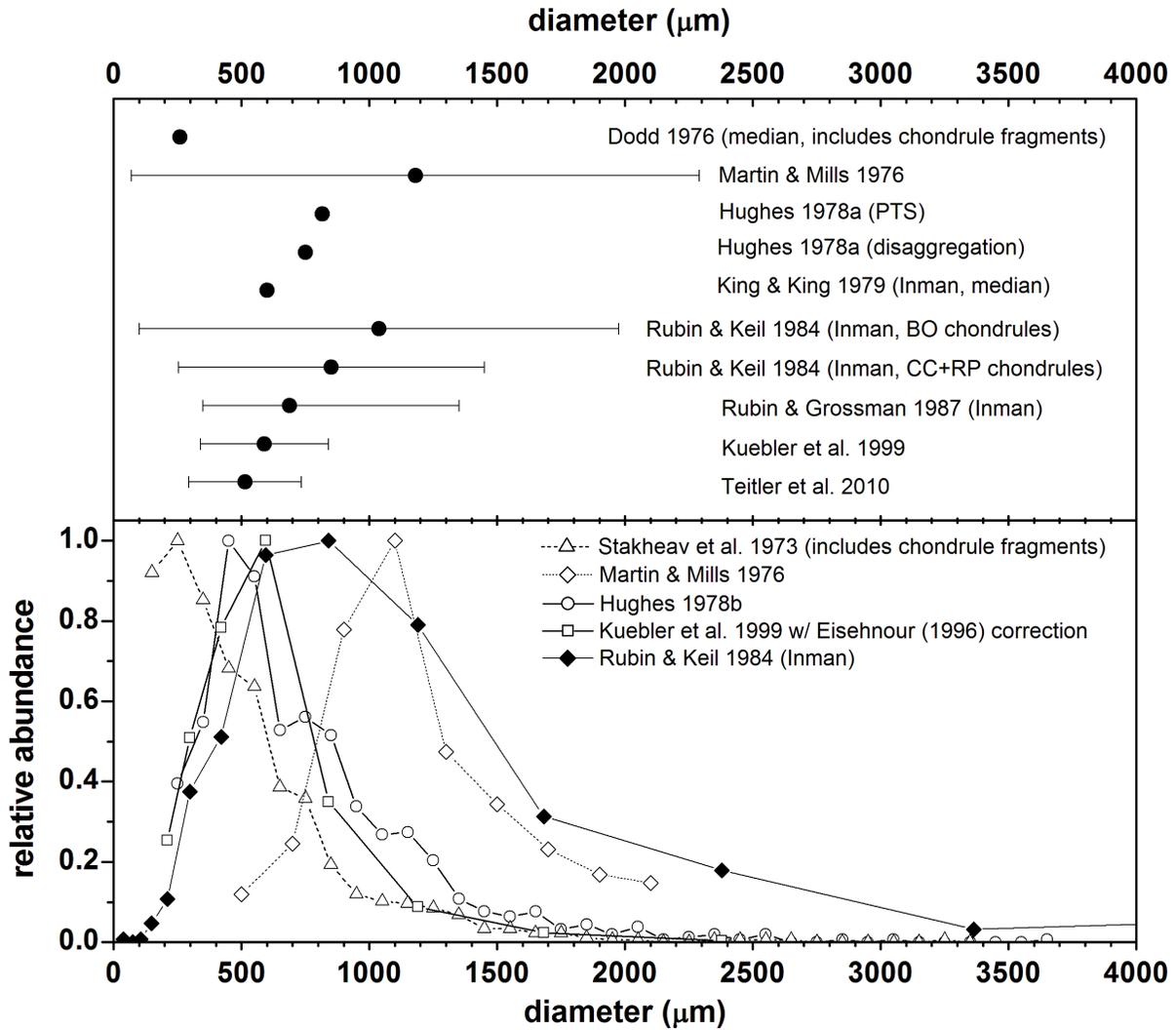

Fig. 4. LL chondrite chondrule mean diameters and chondrule size-frequency distributions. Abscissa has the same scale as Figs. 1-3,5. The LL chondrites are the most extensively studied among the OCs and have the best constrained mean and size frequency distribution. See Table 4 for numerical data and methodology notes.

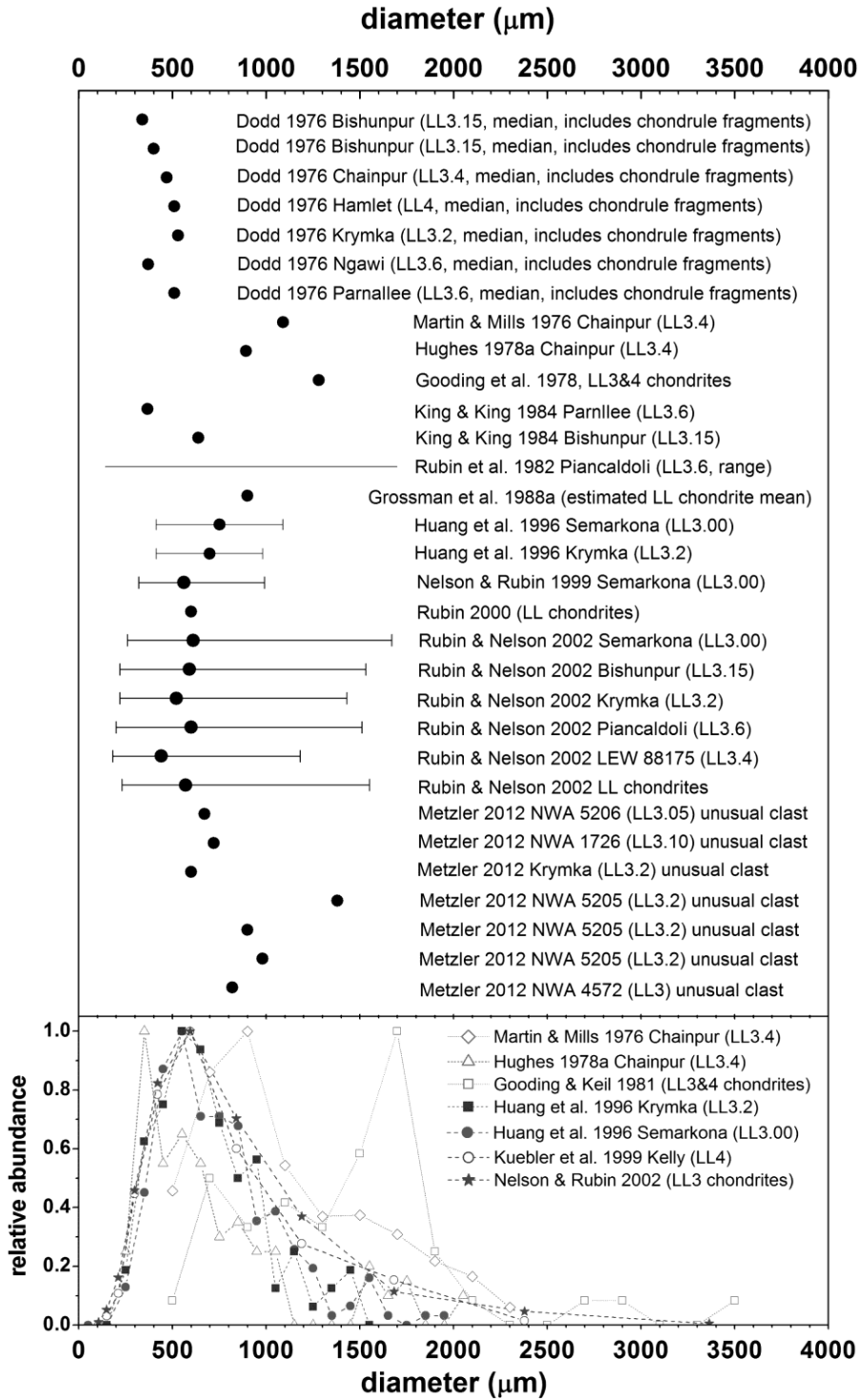

Fig. 5. Enstatite (EL and EH), R and K chondrite chondrule diameters and size-frequency distributions. Abscissa has the same scale as Figs. 1-4. EH chondrite chondrules are generally about half the diameter of EL and OC chondrules, while R and K chondrite chondrules are similar in diameter to OC chondrules. See Table 5 for data and commentary.

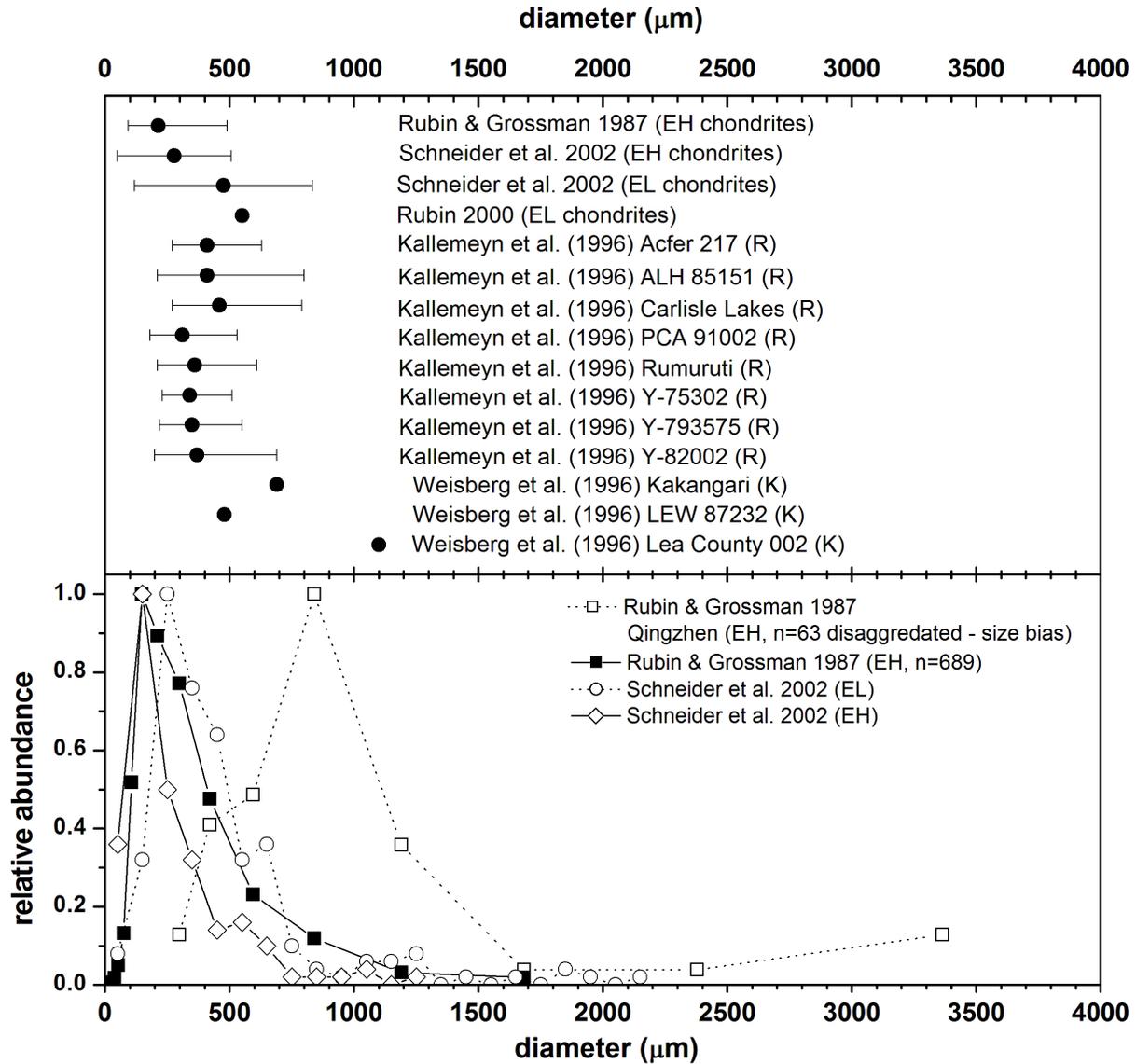

Fig. 6. Reported carbonaceous chondrite chondrule diameters and the size-frequency distribution of CO chondrules. See Table 6 for related numerical data and commentary. Note that the abscissa scale is different than other figures in this compilation.

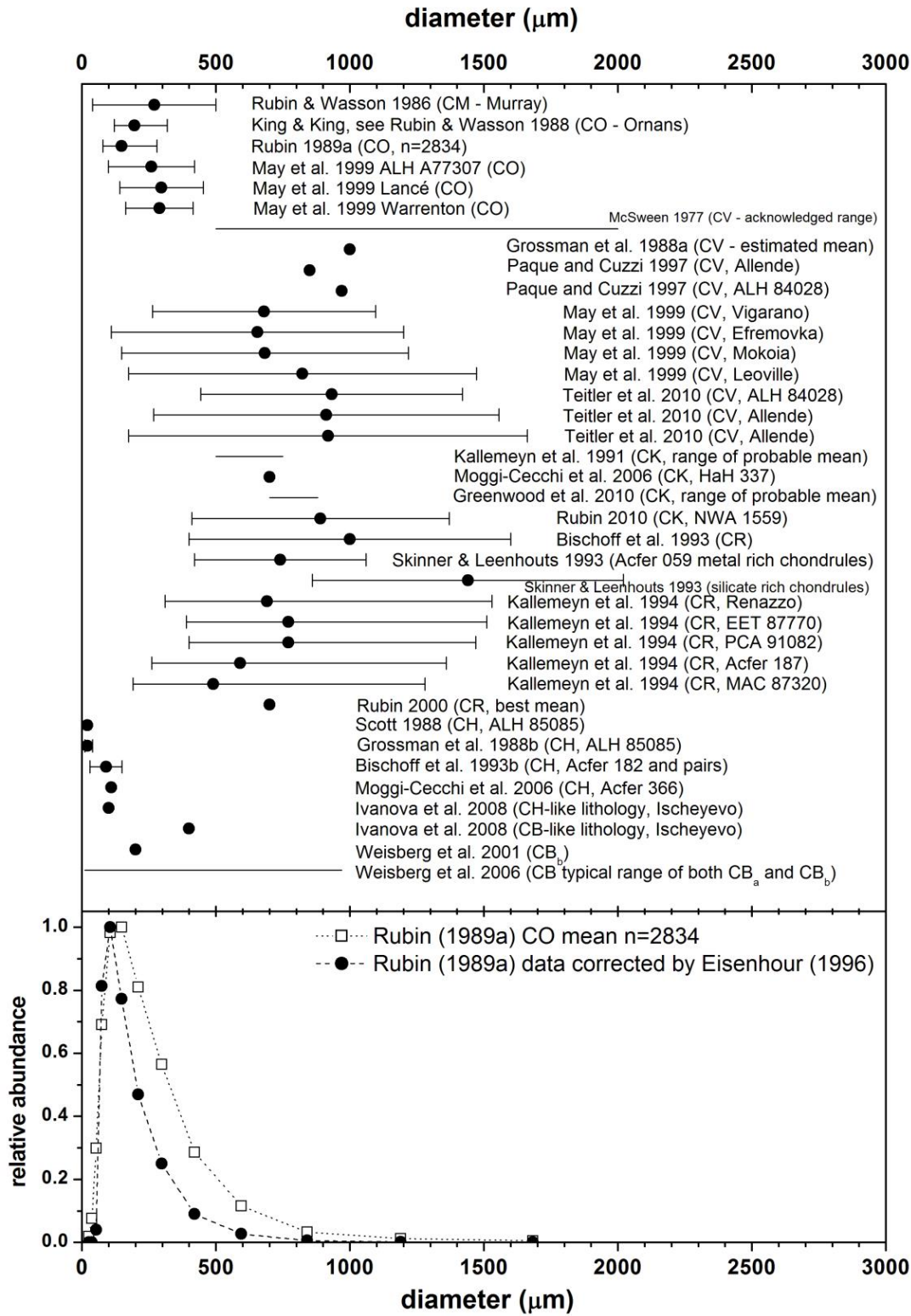

Fig. 7. Comparison of size-frequency distributions of ordinary chondrite chondrules. The ordinary chondrites possess very similar mean chondrule diameters: H (~450 µm) – L (~500 µm) – LL (~550 µm), a result of the positive (coarser) skewness of each group's distribution increasing H<L<LL.  Hence, when a mean (assuming a log-normal distribution) is calculated, mean chondrule diameters increase H<L<LL.  It is unknown if the increasing skewness reflects (unmelted) precursor size or another or another astrophysical parameter such as increased chondrule recycling in the LL chondrites relative to the H chondrites.

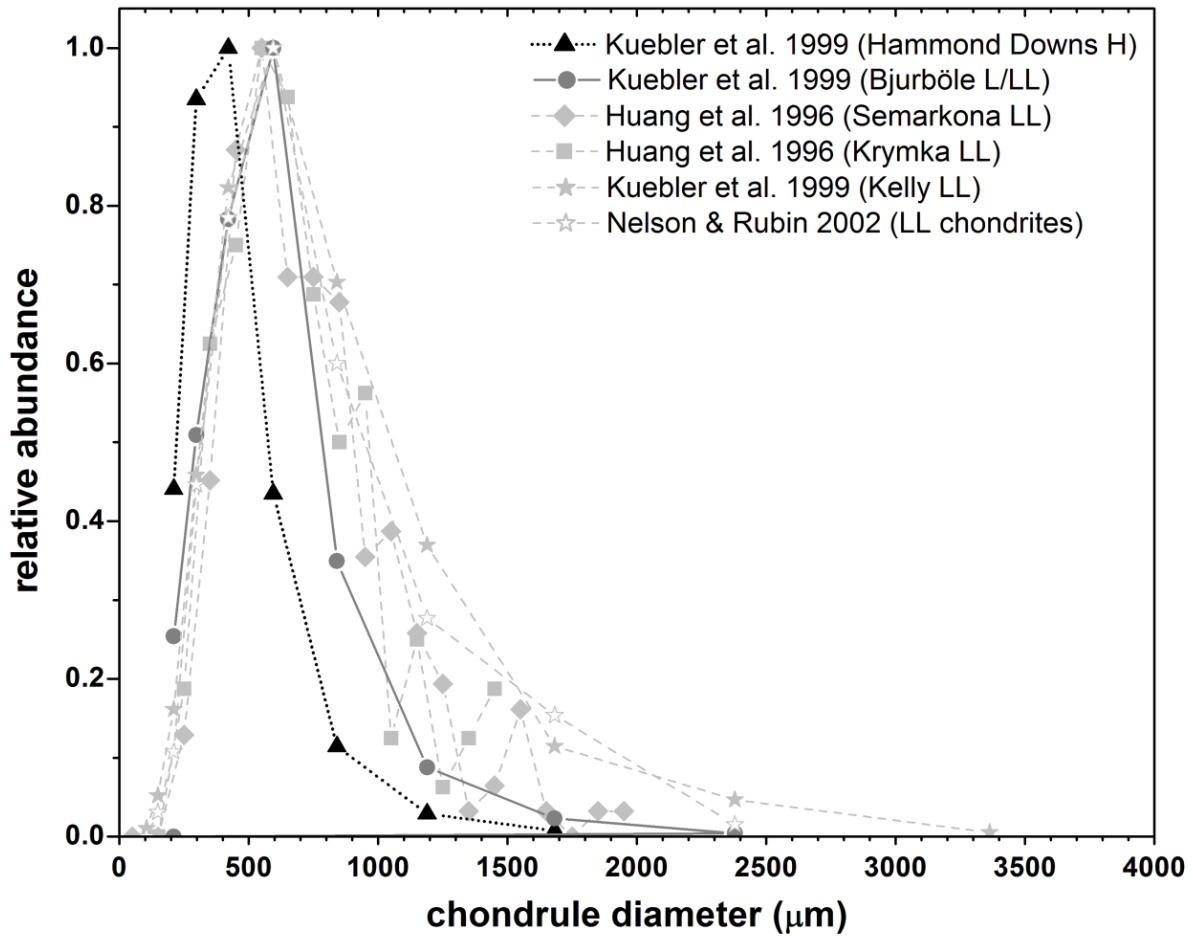